\documentclass[10pt]{llncs}
\pdfoutput=1
\usepackage[utf8]{inputenc}
\usepackage[T1]{fontenc}

\usepackage[margin=0.925in]{geometry}
\usepackage{amsmath,amsfonts, amssymb}
\usepackage{bbm}
\usepackage{mathtools, thmtools}
\usepackage{xfrac}
\usepackage{indentfirst}
\usepackage{braket}
\usepackage{xcolor}
\usepackage[colorlinks,bookmarks=true]{hyperref}
\usepackage[capitalise,nameinlink]{cleveref}
\usepackage[shortlabels]{enumitem}
\usepackage{svg}
\usepackage{subcaption}
\usepackage{cancel}
\usepackage{array, booktabs, makecell}
\usepackage[normalem]{ulem}
\usepackage{tabularx, float} 
\usepackage{caption}
\usepackage{authblk}
\usepackage{hhline}
\usepackage{framed}
\usepackage{multicol}

\usepackage[backend=biber,
style=alphabetic,backref=true,maxnames=4,minnames=3,maxbibnames=99]{biblatex}
\DefineBibliographyStrings{english}{
  backrefpage={Cited on page},
  backrefpages={Cited on pages}
}

\setenumerate[1]{wide = 0pt,labelwidth = 0em, itemsep = 0em, leftmargin=*}
\newlist{defaultenumerate}{enumerate}{3}
\setlist[defaultenumerate,1]{label=\arabic*.}%
\setlist[defaultenumerate,2]{label=\arabic*.}%
\setlist[defaultenumerate,3]{label=\arabic*.}%

\allowdisplaybreaks[1]

\newcommand{\calA}{\mathcal{A}}
\newcommand{\calS}{\mathcal{S}}
\newcommand{\calB}{\mathcal{B}}
\newcommand{\calC}{\mathcal{C}}

\newcommand{\UFAlg}{\mathcal{A}}
\newcommand{\SRAlg}{\mathcal{B}}
\newcommand{\MRAlg}{\mathcal{S}}
\newcommand{\TDAlg}{\mathcal{C}}
\newcommand{\ROAlg}{\mathcal{D}}
\newcommand{\RTAlg}{\mathcal{E}}
\newcommand{\ISAlg}{\mathcal{F}}

\newcommand{\Gufcma}{G_{ufcma}}
\newcommand{\Gsignrep}{G_{signs}}
\newcommand{\Gmeasrep}{G_{c. queries}}
\newcommand{\Gtrapd}{G_{trapdoor}}
\newcommand{\GRepOne}{G_{b1-prime}}
\newcommand{\GRepTwo}{G_{b1-doub-prime}}
\newcommand{\GRISIS}{G_{risis}}

\newcommand{\calN}{\mathcal{N}}
\newcommand{\Adv}{\mathrm{Adv}}
\newcommand{\Func}{\mathrm{Func}}
\newcommand{\LWE}{\mathsf{LWE}}
\newcommand{\SVP}{\mathsf{SVP}}
\newcommand{\SIS}{\mathsf{SIS}}
\newcommand{\RSIS}{\mathsf{R\textnormal{-}SIS}}
\newcommand{\ISIS}{\mathsf{ISIS}}
\newcommand{\RISIS}{\mathsf{R\textnormal{-}ISIS}}
\newcommand{\NTRU}{\mathsf{NTRU}}
\newcommand{\Time}{\mathsf{Time}}

\newcommand{\Z}{\mathbb{Z}}

\newcommand{\N}{\mathbb{N}}
\newcommand{\R}{\mathbb{R}}

\newcommand{\SIG}{\mathsf{SIG}}
\newcommand{\KeyGen}{\mathsf{KeyGen}}
\newcommand{\Sign}{\mathsf{Sign}}
\newcommand{\Verify}{\mathsf{Verify}}
\newcommand{\FALCON}{\mathrm{FALCON}}
\newcommand{\Klein}{\mathrm{KleinS}}
\newcommand{\UFNMA}{\mathsf{EUF\textnormal{-}NMA}}
\newcommand{\UFCMA}{\mathsf{EUF\textnormal{-}CMA}}
\newcommand{\sUFCMA}{\mathsf{sEUF\textnormal{-}CMA}}
\newcommand{\sUFNMA}{\mathsf{sEUF\textnormal{-}NMA}}
\newcommand{\BKZ}{\mathsf{BKZ}}

\let\oldforall\forall
\let\forall\undefined
\let\oldexists\exists
\let\exists\undefined
\DeclareMathOperator{\exists}{\, \oldexists \,}
\DeclareMathOperator{\forall}{\oldforall \,}

\newcommand{\abs}[1]{| #1 |}
\newcommand{\sample}{\overset{{\scriptscriptstyle\$}}{\leftarrow}}
\newcommand{\ketbra}[2]{\mathinner{|{#1}\rangle\hspace{-1pt}\langle{#2}|}}
\newcommand{\ketbrasame}[1]{\ketbra{#1}{#1}}

\newcommand{\negl}{\mathrm{negl}(\eta)}

\newcommand{\Rot}{\mathsf{Rot}}
\newcommand{\gs}[1]{|| #1 ||_{GS}}

\begin{document}
\title{Bounding the security of the FALCON signature scheme in the quantum random oracle model}

\author{Kelsey A. ~Jackson\inst{1}\orcidID{0000-0002-7586-6111}\thanks{kaj22475@UMD.edu} \and Carl A. ~Miller\inst{1, 2}\orcidID{0000-0003-1917-1531}\thanks{camiller@umd.edu}}
\institute{Joint Center for Quantum Information and Computer Science (QuICS), University of Maryland \and Computer Security Division, National Institute of Standards and Technology (NIST)}

\date{\vspace{-6ex}}
\maketitle

\begin{abstract}
In 2016, the National Institute of Standards and Technology (NIST) announced their invitation for proposals of public key encryption and signature schemes that are resistant to attacks by quantum computers. Since then, several rounds of proposals and revisions have occurred resulting in the 2022 announcement that three signature schemes - CRYSTALS-Dilithium, $\FALCON$, and SPHINCS+ - would be standardized. However, in 2024, work by Fouque et al. noted an oversight in the published security for $\FALCON$, which they were only able to remedy in a classical computing setting, namely the random oracle model (ROM). In this paper, we provide the first complete security proof for $\FALCON$ in the quantum random oracle model (QROM). Specifically, we will prove a reduction to the security of $\FALCON$ from the ring inhomogeneous short integer solution ($\RISIS$) problem and the $\NTRU$ learning problem utilizing known quantum reprogramming results. We then provide parameter sets which meet the security levels required by NIST under our reduction. \\
\end{abstract}


\section{Introduction}
Quantum computers are able, in theory, to perform certain tasks much faster than classical computers. These tasks include factoring and the discrete logarithm problem, two of the mathematical problems that underlie current protocols for public-key cryptography. As quantum computers become more powerful and readily available, new cryptographic protocols and primitives must be designed to resist quantum attacks. To this end, NIST's Post-Quantum Cryptography (PQC) Standardization Project aims to invite and review proposals for digital signature schemes and public key encryption protocols that are thought to be secure against adversaries with access to quantum resources. In 2022, NIST selected three digital signature schemes for official standardization: CRYSTALS-Dilithium (ML-DSA), $\FALCON$ (FN-DSA), and SPHINCS+ (SH-DSA). These schemes were selected in recognition of their compactness, efficiency, and the robustness of their security proofs \cite{NIST25}. However, a 2024 review of the security proof given for $\FALCON$ suggested that further evaluation is needed \cite{CloserLook}. 
The authors of \cite{CloserLook} found gaps in the security analysis outlined in the original FALCON specification \cite{FALCON} and they gave a security proof that filled in those gaps. However, their proof was only in the classical random oracle model (ROM) rather than in the quantum random oracle model (QROM).

The classical ROM assumes that any hash function $F: X\rightarrow Y$ used in the digital signature scheme can be replaced with a function selected uniformly at random from the set of all functions mapping $X$ to $Y$. This is a standard assumption in cryptosystems that use hash functions, and it is theoretically beneficial in that it is easier to work with random functions rather than real hash functions.

However, the ROM is limited because it does not account for the possibility that the adversary attacking the scheme could apply the hash function to a superposition of input strings. To generalize to a quantum setting, we need to consider the QROM. For a randomly selected function $H$ mapping $X$ to $Y$, we define an isometry,
\begin{align}
    H_Q:\sum_{x \in X} \alpha_x \ket{x} \rightarrow \sum_{x \in X} \alpha_x \ket{x} \ket{H(x)}.
\end{align}
The QROM assumes any hash function from $X$ to $Y$ can be replaced with a black-box that accepts a quantum state supported on $X$ and returns a quantum state over $X\times Y$ which is computed by some truly random function from $X$ to $Y$.

The goals of this paper are to (i) confirm the post-quantum security of $\FALCON$ by providing a new security proof in the quantum random oracle model (QROM) and (ii) provide compatible parameters that can be proven to reach the concrete security levels requested by NIST as part of the PQC project. Our security proof combines an adaptive reprogramming technique for quantum oracles with a statistical analysis of the smoothing properties of NTRU-based trapdoor sampling \cite{Reprogram, RandomOracles}.

\subsection{The $\FALCON$ signature scheme}\label{sec:FALCON}
$\FALCON$ \cite{FALCON} is a lattice-based signature scheme that, in order to guarantee a smaller key size, restricts its lattice problems to being over NTRU lattices, that is, lattices constructed as follows.
Let $q$ be a prime, $n$ a power of $2$, and let $\phi = (x^n + 1 ) \in \mathbb{Z}[x]$, and $R_{q,n} = \mathbb{Z}_q[x]/(\phi)$.
Let $f,g,F,G \in \mathbb{Z}[x]/(\phi)$ be polynomials with small coefficients such that
\begin{align}
    fG - gF = q \mod \phi,
\end{align}
and such that $f$ is invertible when considered as an element of $R_{q,n}$. 
Let $h \coloneqq gf^{-1}\mod q$. Then, the two matrices
\begin{align}
    &\left[\begin{array}{c|c}
        1 &h \\ \hline
        0 &q
    \end{array}\right],
    &\left[\begin{array}{c|c}
        f &g \\ \hline
        F &G
    \end{array}\right]
\end{align}
generate the same lattice in $\mathbb{Z}^{2n}$. FALCON derives its public key from the first matrix above and its secret key from the second.
Security of FALCON is based in part on the assumption that it is computationally hard, for a given polynomial $h \in R_{q, n}$, to compute small polynomials $f,g$ such that $h = g f^{-1} \mod q$.

$\FALCON$ utilizes a norm and an inner product $\langle \cdot , \cdot \rangle: \Z[x]/(\phi)\times \Z[x]/(\phi) \rightarrow \mathbb{C}$ defined as follows: for $a, b \in \Z[x]/(\phi)$.
\begin{align}
\langle a, b\rangle \coloneqq \frac{1}{\deg({\phi})}\sum_{\phi(\zeta)=0}a(\zeta)b(\zeta)^*, \qquad \abs{a} \coloneqq \sqrt{\langle a, a\rangle},
\end{align}
where $\zeta$ varies over the roots of $\phi$. Then, for $\phi = x^n+1$ and $n$ a power of $2$, we may also write $\langle a, b\rangle = \sum_{i=0}^{n-1}a_i b_i$ for $a_i, b_i$ the $i$th coefficients of $a, b$, respectively. Now for vectors $u, v \in \left(\Z[x]/(\phi)\right)^m$ and $m \in \N$ we define $\langle u, v \rangle \coloneqq \sum_{j=0}^{m-1} \langle u_j, v_j\rangle.$ 

Let $D_{\Lambda, \sigma, \mu}$ denote a discrete Gaussian distribution over a lattice $\Lambda$ with mean $\mu$ and standard deviation $\sigma$. Let $\Klein(B, \sigma, t)$ denote the sampling algorithm given in Figure \ref{fig:Klein}. This algorithm, given some lattice basis $B$, variance $\sigma^2$, and target vector $t$, returns a vector $z$ such that $Bz$ is statistically indistinguishable from a vector selected randomly from $D_{\Lambda, \sigma, t}$ where $\Lambda$ is the lattice generated by the columns of $B$ within our parameter regime \cite{FALCON, Klein}.

\begin{figure}[H]
\centering
\begin{framed}
\begin{center} 
\noindent \underline{$\Klein(B, \sigma, t)$}
\begin{enumerate}
    \item Let $\tilde{B}$ be the Gram-Schmidt orthogonalization of $B$ such that for some upper-triangular matrix $U$, $B = \tilde{B}\cdot U$. Let $\tilde{b}_i$ be the $i$th column vector in $\tilde{B}$
    \item $\sigma_j \coloneqq \sigma/\abs{\tilde{b}_j} \forall j \in [m]$
    \item For $j = m, m-1, ..., 1$:
    \begin{enumerate}
        \item $c_j \coloneqq t_j + \sum_{i>j}(t_i-z_i)U_{i, j}$
        \item $z_j \leftarrow D_{\Z, \sigma_j, c_j}$
    \end{enumerate}
    \item Return $z$
\end{enumerate}
\end{center}
\end{framed}
\caption{Klein's sampling algorithm}
\label{fig:Klein}
\end{figure}

We now give a brief overview of the hash-and-sign signature scheme, $\FALCON$. Let $\lambda \coloneqq (q, n, k, \sigma, \sigma_{f, g}, \beta)$ be parameters and let $H: R_{q, n} \times \{0, 1\}^k\times\{0, 1\}^* \rightarrow R_{q, n}$ be a hash function. Figure \ref{fig:falcon_simplified} describes a simplified version of the $\FALCON$ protocol.
\begin{figure}[H]
\centering
\begin{framed}
\begin{center} 
\textbf{The $\FALCON$ protocol (simplified)} 
\begin{multicols}{2}
    \noindent \underline{$\KeyGen(\lambda)$}
\begin{enumerate}
    \item $f_i, g_i \sample D_{\Z, \sigma_{f, g}, 0} \forall i \in [n]$
    \item $f \coloneqq \sum_{i\in[n]} f_i x^i$
    \item $g \coloneqq \sum_{i\in[n]} g_i x^i$
    \item Compute $F, G \ni fG -gF = q\mod \phi$
    \item $B \coloneqq \begin{bmatrix}g &-f \\ G &-F\end{bmatrix}$
    \item $(pk, sk) \coloneqq (h\coloneqq g f^{-1} \mod q, B)$
\end{enumerate}
    \noindent \underline{$\Verify(\lambda, pk, M, (r, s_2))$}
\begin{enumerate}
    \item $c\coloneqq H(pk, r, M)$
    \item $s_1 \coloneqq c-s_2 h \mod q$
    \item Return $\vert (s_1, s_2) \vert^2 \le \beta^2$
\end{enumerate}
    \pagebreak
    \noindent \underline{$\Sign_H(\lambda, sk, M)$}
\begin{enumerate}
    \item $r\sample \{0, 1\}^{k}$
    \item $c \coloneqq H(pk, r, M)$
    \item $s_2 \coloneqq \bot$
    \item While $s_2 = \bot$:
    \begin{enumerate}
        \item $(s_1, s_2) \leftarrow \Klein(B, \sigma, (c, 0))$
        \item If $\vert (s_1, s_2) \vert^2 > \beta^2$, $s_2 \coloneqq \bot$.
    \end{enumerate}
    \item Return $(r, s_2)$
\end{enumerate}
    \noindent \underline{$\Sign^+_H(\lambda, sk, M)$}
\begin{enumerate}
    \item $s_2 \coloneqq \bot$
    \item While $s_2 = \bot$:
    \begin{enumerate}
        \item $r\sample \{0, 1\}^{k}$
        \item $c \coloneqq H(pk, r, M)$
        \item $(s_1, s_2) \leftarrow \Klein(B, \sigma, (c, 0))$
        \item If $\vert (s_1, s_2) \vert^2 > \beta^2$, $s_2 \coloneqq \bot$.
    \end{enumerate}
    \item Return $(r, s_2)$.
\end{enumerate}
\end{multicols}
\end{center}
\end{framed}
\caption{A simplified description of the key generation algorithm ($\KeyGen$), signature verification algorithm ($\Verify$) and signing algorithm ($\Sign_H$) for $\FALCON$ \cite{FALCON}. The modified $\FALCON$ scheme proposed in "A Closer Look..." replaces $\Sign_H$ with $\Sign^+_H$ and is named $\FALCON^+$ \cite{CloserLook}.}
\label{fig:falcon_simplified}
\end{figure}
Note that the choice of $f, g$ determine $F, G$ albeit at the cost of a relatively expensive computation to solve the NTRU condition. Therefore, for storage purposes, $F$ is generally stored alongside $f, g$ to allow for a faster calculation of $G$ alone \cite{FALCON}. As shorthand, we may write $\Verify(M, \sigma)$ to indicate $\Verify(\lambda, pk, M, \sigma)$ when the parameters $\lambda$ and public key $pk$ are not relevant to the argument at hand. The same applies to $\KeyGen$ and $\Sign_H$ where applicable.

\paragraph{$\FALCON$ vs. $\FALCON^+$:} $\FALCON$ as originally designed, utilizes $\Sign_H$ as described in Figure \ref{fig:falcon_simplified}. This algorithm, given a secret key and message, makes a single call to the hash function at $r \sample \{0, 1\}^k$ then allows the $\Klein$ sampling algorithm to make an arbitrary number of attempts to create a valid signature utilizing this query. This works because $\Klein$ is a randomized algorithm that can be expected to succeed with high probability after some constant number of attempts (generally one or two repetitions for the proposed parameters \cite{CloserLook}). This choice is logical for practical purposes, but a relatively unusual choice theoretically compared to more traditional Hash-and-Sign designs. Instead, as proposed by the Fouque et al. result, we considered a variation on $\FALCON$ nicknamed $\FALCON^+$, which makes a new query to the hash function for every attempt by $\Klein$ to create a valid signature. Ultimately, however, it proved useful to explore this modification in greater depth which allowed us to show that the two signing oracles are statistically indistinguishable under the parameter regime considered. Our security result will therefore apply to both $\FALCON^+$ and $\FALCON$. This is useful as, in response to the Fouque et al. results, this modification is planned to be integrated into future implementations of $\FALCON$ as well as the resulting FIPS standard regardless \cite{CloserLook}. Therefore, our analysis can be applied directly to the $\FALCON$ specification published as of the time of writing, but it is also expected to apply to future iterations of the scheme. For this document, we will use $\FALCON$ and $\FALCON^+$ as specified in Figure \ref{fig:falcon_simplified} and $\FALCON^{(+)}$ when discussing both.

\subsection{Prior security proofs for FALCON}
Security levels for digital signatures can be quantified in terms of existential unforgeability against no-message attacks ($\UFNMA$) and existential unforgeability against chosen-message attacks ($\UFCMA$). For the latter, adversaries are allowed to make sequential queries to some black-box signing oracle and then attempt to forge a valid message and signature pair other than those retrieved from the oracle. Our main security results use the $\UFCMA$ model.

As originally published in the proposed specification for $\FALCON$, the security of the scheme was based on applying the Gentry, Peikert and Vaikuntanathan (GPV) framework, which allows for secure hash-and-sign signature schemes using pre-image sampleable trapdoor hash functions relying on the hardness of the Short Integer Solution ($\SIS$) problem \cite{GPV}. $\FALCON$'s specification then proposes that adjusting this result to the more structured case of $\SIS$ over NTRU lattices is straightforward. Their analysis concluded that the best key-recovery attack would require a full reduction of the lattice while the best signature-forgery attack would require solving the $\gamma-\SVP$ problem, for which the currently optimal attack also relies on lattice reduction \cite{FALCON}. If accurate, this argument would result in a security proof for the $\UFCMA$ security of $\FALCON$ based on the hardness of $\SIS$ and $\gamma-\SVP$ over NTRU lattices.

However, as more deeply discussed in a recent result from Fouque et al., adapting the GPV framework to NTRU lattices over rings within the parameter regime desired by $\FALCON$'s authors is substantially more complicated than initially thought. This occurs for two reasons. First, $\FALCON$'s parameter calculations focus on bounding the Rényi divergence to quantify the statistical nearness of their pre-image sampler to $D_{\Z, \sigma, 0}$, rather than the more conventional statistical distance. Second, the GPV framework implicitly utilizes the Leftover Hash Lemma to argue that for an $\SIS$ instance $A \in \Z_q^{k\times m}$ and $s \leftarrow D_{\Z, \sigma, \mu}$, a function $H'$ outputting $A\cdot s \mod q$ is practically indistinguishable from a random oracle. This result does not directly apply to the ring structure used in $\FALCON$ and if it is adapted naively, then we find the output of $H'$ to have statistical distances from uniform as high as $2^{-34}$ \cite{GPV, FALCON, CloserLook}. More recent efforts have been made to adapt the leftover hash lemma to rings of varying structures \cite{RingLHL}. However, in $\FALCON$'s parameter regime with $\sigma_{f, g} \approx 1.17\sqrt{q/2n}$, applying the former results without adaptation does not improve the bound on the statistical distances compared to those given by Fouque et al. \cite{RingLHL, CloserLook}.

The authors follow this with reductions from the $\RSIS$ problem to the strong unforgeability of $\FALCON^+$, and from a multi-target inhomogeneous $\RSIS$ ($t-\RISIS$) problem to the plain unforgeability. However, a major caveat for this analysis is that it models all queries made to the random oracle as classical, and relies on this trait while accounting for any instances of repeated queries \cite{CloserLook}. This is not a behavior that can be abstracted to a quantum setting where it is both possible and common for algorithms to make repeated queries consisting of superpositions over the domain. As a result, the Fouque et al. approach cannot directly address the post-quantum security of $\FALCON^{(+)}$ \cite{CloserLook}. Therefore, in order for $\FALCON^{(+)}$ to truly function as a post-quantum cryptographic scheme, it is necessary to provide a new proof of the security of the scheme in the QROM.

\subsection{Overview of main results}
In order to provide an $\UFCMA$ security proof for $\FALCON^+$ that is functional in the QROM, we introduce a series of security games. We prove that an efficient algorithm for any game in the sequence implies an efficient algorithm for the next game in the sequence, without too considerable of a loss in success probability. The first game in the sequence is the $\UFCMA$ experiment for $\FALCON^+$, and the last game involves solving a form of the $\RISIS$ problem. A key part of this series of games is the application of the Measure \& Reprogram technique from \cite{Reprogram}, which converts a quantum oracle algorithm into an algorithm for a $2$-round interactive protocol.
The series of games can be characterized as follows:

\begin{enumerate}
    \item Game $\Gufcma$: The $\UFCMA$ experiment for $\FALCON^+$. A quantum algorithm $\calA$ is allowed $Q$ quantum queries to $H$ and $Q$ classical queries to $\Sign^+_H$, and attempts to forge a new signature.
    \item Game $\Gsignrep$: The $\UFCMA$ experiment for $\FALCON$. This replaces $\Sign^+_H$ with $\Sign_H$ and defines the latter's behavior as $B_2O^HB_1$ for unitaries $B_1, B_2$ and $O^{\ket{H}}$ the query to $H$.
    \item Game $\Gmeasrep$: This is the game that arises from Game $\Gsignrep$ after the application of the Measure \& Reprogram technique from \cite{Reprogram}.
    \item Game $\Gtrapd$: We replace $H$ with a simulated random oracle containing a family of trapdoor functions that --- given an NTRU public key $h_B$ --- outputs samples of the form $[1\mid h_B]\cdot(s_1,s_2)^T \mod q$, where $s_1,s_2$ are independently sampled from a discrete Gaussian distribution over $R_{q,n}$.
    \item Game $\GRepOne$: We modify $B_1, B_2$ to less closely rely on use of the secret key $B$.
    \item Game $\GRepTwo$: We modify $B_1, B_2$ to be independent of the secret key $B$.
    \item Game $\GRISIS$: The Hermite-form $\RISIS$ problem.
\end{enumerate}

The first step (replacing $\Sign^+_H$ with $\Sign_H$) is a simple but helpful simplification, as this modification then permits the application of the Measure \& Reprogram Lemma (Lemma \ref{lem:measure_and_reprogram}). We then decouple our adversary's access to $\Sign_H$ from any access to the secret key $B$. Once all of this is accomplished, we recognize that to forge a signature, the adversary $\calA$ must solve the following:
\begin{align}
    \Pr_{h, t\sample R_{q, n}}\Bigl[[1|h]\cdot y = t \mod q \wedge \abs{y}^2\le \beta^2\mid y \leftarrow\calA(h, t) \Bigr].
\end{align}
This is equivalent to the Hermite-form $\RISIS$ problem. Thus, we conclude that for any polynomial quantum query algorithm $\calA$ there exist efficient algorithms $\calN, \ISAlg$ such that:
\begin{align}
    \Adv^{Q-\UFCMA}_{\FALCON^{+}}(\calA) \le (4Q+1)^2\left(\Adv^{\NTRU}_{\sigma_{f, g}}(\calN) + \Adv^{\RISIS}_{1, \beta}(\ISAlg)\right) + \negl
\end{align}
for $Q$ the number of queries and $\eta$ a tuple of other parameters. Then, our analysis is instantiated using an $\NTRU$-based trapdoor construction, with the resulting security guarantee relying on the hardness of the $\NTRU$ learning problem:
\begin{align}
    \Adv^{\NTRU}_{\sigma_{f, g}}(\calN) \coloneqq \left| \Pr[b=1|f, g \leftarrow \mathbb{D}_{R_{q, n}, \sigma_{f, g}, 0}, h_B\coloneqq gf^{-1}\mod q, b\leftarrow \calN(h_B)] \right. \\ \left. - \Pr[b=1|h_B\sample R_{q, n}, b\leftarrow \calN(h_B)] \right|,
\end{align}
where $f, g$ are selected from a discrete Gaussian distribution over $R_{q, n}$ with standard deviation $\sigma_{f, g}$ centered on $0$. Finally, we define the advantage of an algorithm $\ISAlg$ to solve the inhomogeneous short integer solution problem over the ring $R_{q, n}$, also known as $\RISIS$ (Hermite form):
\begin{align}
    \Adv^{\RISIS}_{1,\beta}(\ISAlg) \coloneqq \Pr\Bigl[[1|A]\cdot y = t \mod q \wedge \abs{y}^2\le \beta^2\mid A\sample R_{q, n}, t\sample R_{q, n}^1, \ y \leftarrow\calA(A, t) \Bigr].
\end{align}

We now consider the sources of loss and possible future improvements to our reduction. The dominant $O(Q^2)$ loss arises from the adaptive reprogramming step; quadratic loss for adaptive reprogramming is known to be unavoidable in general, and we further establish a corresponding optimality result for hash-and-sign schemes in Section \ref{sec:tightness}. Although the particular structure of this scheme may permit a tighter reprogramming argument, we do not expect a substantial improvement from this step. The dependence on $\mathsf{Adv}(\NTRU)$ arises from the way our reduction relates an intermediate game to the underlying $\NTRU$ problem. Therefore, a different approach could potentially improve the multiplicative factors accompanying this term, but removing it entirely would likely require a fundamentally different reduction. Remaining losses arise from the choice of sampling algorithm and from statistical approximations used in the reduction, including Bernoulli's inequality, though this is well-controlled in the parameter regime we consider. Regardless, improvements to these components are expected to have a lesser effect on the resulting bound than the loss introduced by adaptive reprogramming. Thus, while some factors may be improved, the $O(Q^2)$ loss from adaptive reprogramming appears to be the primary limitation on the tightness of our result.
\section{Preliminaries}\label{sec:Prel}
Let $\N$ denote the set of positive integers. Then, for $k \in \N$, $[k] \coloneqq \{1, 2, ..., k\}$. An alphabet is a finite non-empty set. Given an alphabet $S$, the notation $s \sample S$ indicates an element $s$ selected uniformly at random from $S$. Given two alphabets $A, B$, the notation $\Func(A, B)$ indicates the set of all functions mapping $A$ to $B$. For arbitrary strings $a, b$ we express the concatenation of $a, b$ as $a \vert\vert b$. Similarly, given matrices $A_1, \dots A_k$ with the same number of rows, $\left[A_1|\dots|A_k\right]$ denotes the matrix consisting of the $A_i$ matrices placed side by side. The expression $\log$ indicates the base-$2$ logarithm. 
We use $\ni$ as shorthand for "such that". 

Throughout this paper, $n$ denotes a power of $2$, and $q$ denotes a prime number such that $2n \mid (q-1)$. Let $\abs{\cdot}$ denote the Euclidean norm unless otherwise specified.

We let $R_{q, n}$ denote the
ring $\mathbb{Z}_q [ X ] / (X^n + 1 )$. The condition that $2n \mid (q-1)$ implies that the polynomial $(X^n + 1 )$ has $n$ distinct roots in $\mathbb{Z}_q$, and therefore 
$R_{q,n} \cong \mathbb{Z}_q^n$.
Note that any basic operations written below that are completed over $R_{q, n}$ should be understood as completed modulo $q$ whether or not $\textnormal{mod } q$ is explicitly written.

For $\mu, \sigma \in \mathbb{R}$ with $\sigma > 0$, let $\rho_{\mu, \sigma} \colon \mathbb{R} \to \mathbb{R}$ be defined by:
\begin{align}
    \rho_{\sigma, \mu}(x) = \exp\left(\frac{-\vert x-\mu\vert^2}{2\sigma^2}\right).
\end{align}
Then, $D_{\Lambda, \sigma, \mu} \colon \Lambda \to \mathbb{R}$ denotes the probability distribution \begin{align}
    D_{\Lambda, \sigma, \mu}(x) = \frac{\rho_{\sigma, \mu}(x)}{\sum_{z\in\Lambda}\rho_{\sigma, \mu}(z)},
\end{align}
which is the discrete Gaussian distribution over the lattice $\Lambda$ with center $\mu$ and Gaussian parameter $\sigma$.

Let $X, Y$ be probability distributions on a set $D$ with $\mathsf{Supp}(X) \subseteq \mathsf{Supp}(Y)$. Let $a \in (1, \infty)$. Then, we define the $a$-Rényi divergence $\mathcal{R}_a$ of $X, Y$ as follows:
\begin{align}
    \mathcal{R}_a(X, Y) \coloneqq \left(\sum_{x\in \mathsf{Supp}(X)}\frac{X(x)^a}{Y(x)^{a-1}}\right)^{1/(a-1)},
\end{align}

Alternatively, we can examine the statistical distance between these same distributions. The statistical distance between two distributions $X, Y$ over a shared domain $D$, is defined:
\begin{align}
    \Delta(X, Y) \coloneqq \frac{1}{2}\sum_{a\in D}\abs{\Pr[x=a\mid x\sample X]-\Pr[y=a\mid y\sample Y]}
\end{align}

Finally, observe that $\Klein$ - as described above - is a randomized algorithm. For some random seed $u$, we notate a deterministic implementation of $\Klein$ as $\Klein(B, \sigma, t; u)$.

\subsection{Quantum computation model}
A quantum state $\rho$ on $\mathbb{C}^d$ is a positive semi-definite matrix in $\mathbb{C}^{d\times d}$ with trace equal to $1$. A pure quantum state is a quantum state of rank $1$. Any pure state can be uniquely written as $\ketbrasame{\psi}$ for some $\ket{\psi} \in \mathbb{C}^d$, and we may express such a state simply as $\ket{\psi}$ for brevity. A register is either an alphabet or a $k$-tuple ($X_1, X_2, ..., X_k)$ for $k \in \N$ where $X_i \forall i\in [k]$ are alphabets. For the former, if the size of the alphabet is $\abs{\Sigma}$, then we call $\abs{\Sigma}$ the size of the register, and a state on the register is a density matrix on $\mathbb{C}^{\abs{\Sigma}}$. For the latter, if the size of the component alphabets are $\abs{\Sigma_i} \forall i\in [k]$, then we call $\abs{\Sigma_1} \times \dots \times \abs{\Sigma_k}$ the size of the register and a state on the register is a density matrix on $\mathbb{C}^{\abs{\Sigma_1}}\otimes\dots\otimes\mathbb{C}^{\abs{\Sigma_k}}$. We assume that each register \(X\) is equipped with a fixed injective encoding into \(\{0,1\}^n\) for some \(n\). (This encoding allows elements of $X$ to be used as inputs to classical and quantum circuits.)

An $k$-qubit register is a register of the form $W = (Y_1, \dots Y_k)$ with bijections $Y_j \leftrightarrow \{ 0, 1 \}$ for each $j$. An elementary gate on $W$ is a unitary operator on the space $\mathbb{C}^{\left| W \right|}$ that arises from tensoring one of the following three operators:
\begin{align}
\begin{split}
    &T\coloneqq \begin{pmatrix}
       1 & 0
       \\
       0 & e^{i\pi/4}
    \end{pmatrix}, \quad 
    H \coloneqq \frac{1}{\sqrt{2}}\begin{pmatrix}
       1 & 1
       \\
       1 & -1
    \end{pmatrix}, \quad \text{or} \quad
    \text{CNOT} \coloneqq \begin{pmatrix}
        1 & 0 & 0 & 0
        \\
        0 & 1 & 0 & 0
        \\
        0 & 0 & 0 & 1
        \\
        0 & 0 & 1 & 0
    \end{pmatrix},
\end{split}
\end{align}
(where it is assumed that $T$ and $H$ act on a single qubit register from $(Y_1, \ldots, Y_k)$ and CNOT acts on two different qubit registers) with identity operators on all the remaining qubits. 
A quantum algorithm $\mathcal{A}$ acting on $W$ is a sequence of elementary gates on $W$. The number of such gates is called the time complexity of the algorithm, and is denoted by $\Time(\calA)$.
For any algorithm of this kind, we can then define a unitary matrix $U$ associated with $\calA$ that is the product of all the elementary gates in order.

Next we define quantum-query algorithms, which are algorithms that are permitted to make queries to a quantum oracle. In the following definition, the register $X$ serves as the input register for the oracle and the register $Y$ serves as the output register for the oracle. There is also a third register, $Z$, which is a work register used only by the intermediate algorithms and not by the oracle. We assume the existence of an isomorphism $\phi \colon Y \to \mathbb{Z}_{\left| Y \right|}$. The addition operation on $\mathbb{Z}_{\left| Y \right|}$ induces an addition operation on $Y$ (which we denote simply by ``$+$'').

\begin{definition}[Quantum-query algorithm]\label{def:qAlg}
Let $k \in \mathbb{N}$, let $X$, $Y$, and $Z$ be registers, and let $\phi \colon Y \to \mathbb{Z}_{\left| Y \right|}$ be a bijection.
A $k$-quantum query algorithm $\calA$ with input register $X$, output register $Y$, and work register $Z$ is a sequence of $k$ quantum algorithms $A_1, \dots A_k$ each acting on the registers $X, Y, Z$.
\end{definition}

Let $U_i \colon \mathbb{C}^{X \times Y \times Z} \to \mathbb{C}^{X \times Y \times Z}$ be the unitary associated with $A_i$. Assume that the algorithm $\mathcal{A}$ queries an oracle given by a function $H : X \rightarrow Y$. Let $\mathcal{O}^H \colon \mathbb{C}^{X \times Y \times Z} \to \mathbb{C}^{X \times Y \times Z}$ be the mapping given by \begin{align} \ket{x}\ket{y}\ket{z} \rightarrow \ket{x}\ket{y + H(x) }\ket{z} \end{align} 
for all $(x,y,z) \in X \times Y \times Z$.
Then, for any initial pure state $\left| \phi \right> \in \mathbb{C}^{X \times Y \times Z}$, the quantum query algorithm $\mathcal{A}$ maps $\left| \phi \right>$ to the vector
\begin{align} \mathcal{O}^HU_k\dots\mathcal{O}^HU_2\mathcal{O}^HU_1 \left| \phi \right>.
\end{align}
If the quantum algorithm $\mathcal{A}$ makes only classical queries to $H$, that is, every query made by $\mathcal{A}$ is supported on a single element of the alphabet $X$ then we will note this by writing it as $\calA^H$. If the algorithm instead makes any quantum queries, that is the algorithm makes queries that are superpositions supported over multiple elements of $X$, then we will write it $\calA^{\ket{H}}$. Note that a quantum algorithm making classical queries is not necessarily equivalent to a classical algorithm making classical queries.

In addition, recall that in $k-\UFCMA$ experiments, while quantum algorithms make quantum queries to $H$, they only make classical queries to other black-box oracles, such as $\Sign_H$ \cite{RandomOracles}. This is because in a post-quantum cryptography setting, we expect the secret key and signing oracle to be held by a classical party. As a result, $\Sign_H$ is not expected to be able to accept non-classical queries.

\subsection{Cryptographic definitions}
For the descriptions of cryptographic protocols that follow, we will let $\eta \in \mathbb{N}$ denote the security parameter. A function $f \colon \mathbb{N} \to \mathbb{R}$ is called negligible if for all $c > 0$, the function $\left| f \right|$ is eventually smaller than the function $k \mapsto k^{-c}$. The expression $\negl$ is used as a placeholder for negligible functions. For each $\eta$, there is an associated tuple $\iota$ which consists of the publicly available parameters associated with the security parameter $\eta$ and accessible to all participants.

\begin{definition}[Digital signature scheme\label{def:sigScheme}] A digital signature scheme is defined by three algorithms $\SIG = (\KeyGen, \Sign, \Verify)$ such that:
\begin{enumerate}
    \item The key generation algorithm $\KeyGen(\iota)$ outputs a public-key, secret-key pair $(pk,sk)$.
    \item The signing algorithm $\Sign(sk,m)$, which maps any message $m\in \{0, 1\}^*$ to a signature $\sigma$.
    \item The verification algorithm $\Verify(pk,m,\sigma)$ outputs a single bit
\end{enumerate}
We say $\SIG$ has correctness error $\gamma$ if for all possible $(pk,sk)$ produced by $\KeyGen(\iota)$ and all $m\in \{0, 1\}^*$, 
\begin{align}
    \Pr[\Verify(pk, m, \sigma) = 0 \mid \sigma\leftarrow \Sign(sk, m)] \leq \gamma.
\end{align}
\end{definition}

\begin{definition}[$k-\UFCMA$]\label{def:UF} Let $\SIG = (\KeyGen, \Sign_H, \Verify)$ be a signature scheme with public parameters $\iota$. Let $\calA$ be a polynomial-time quantum query algorithm with queried oracle $H$. Then:
\begin{align}
    \Adv_{\SIG}^{\UFCMA}(\calA) \coloneqq& \Pr_{(pk, sk) \leftarrow  \KeyGen}[\Verify(pk, m, \sigma) \wedge m \notin \textup{SignQ} \mid (m, \sigma)\leftarrow\calA^{\ket{H}, \Sign_H(sk, \cdot)}(pk)],
\end{align}
where $\textup{SignQ}$ is the set of queries made by $\calA$ to $\Sign_H(sk, \cdot)$, which $\calA$ accesses as a black-box oracle. For security parameter $\eta \in \mathbb{N}$, we say that $\SIG$ is $k-\UFCMA$-secure if for every efficient $k$-quantum-query algorithm $\calA^{\ket{H}}$, we have $\Adv_{\SIG}^{\UFCMA}(\calA) \leq \negl$. 
\end{definition}

\begin{definition}[$k-\sUFCMA$]\label{def:sUF} Let $\SIG = (\KeyGen, \Sign_H, \Verify)$ be a signature scheme with public parameters $\iota$. Let $\calA$ be a polynomial-time quantum query algorithm with queried oracle $H$. Then:
\begin{align}
    \Adv_{\SIG}^{\sUFCMA}(\calA) \coloneqq& \Pr_{(pk, sk) \leftarrow \KeyGen}[\Verify(pk, m, \sigma) \wedge (m, \sigma) \notin \textup{SignQR} \mid (m, \sigma)\leftarrow\calA^{\ket{H}, \Sign_H(sk, \cdot)}(pk)],
\end{align}
where $\textup{SignQR}$ is the set of queries and responses produced by the interaction between $\calA$ and $\Sign_H$. We say that $\SIG$ is $k-\sUFCMA$-secure if for every polynomial-time (\textit{efficient}) $k$-quantum-query algorithm $\calA^{\ket{H}}$, we have $\Adv_{\SIG}^{\sUFCMA}(\calA) \leq \negl$. 
\end{definition}
Observe that a scheme with $k-\sUFCMA$ security guarantees both $k-\UFCMA$ security and that given a message/signature pair, an adversary cannot produce a new signature for that same message.

\begin{definition}[$\mathsf{(s)}\UFCMA$]\label{def:CMA}
A signature scheme $\SIG = (\KeyGen, \Sign_H, \Verify)$ is $\UFCMA$-secure (resp. $\sUFCMA$-secure) against quantum adversaries if it is $k-\UFCMA$-secure (resp. $k-\sUFCMA$-secure) for every polynomially bounded $k=k(\eta)$ as defined in Definition \ref{def:UF} (resp. \ref{def:sUF}).
\end{definition}

\begin{definition}[$\mathsf{(s)}\UFNMA$]\label{def:NMA}
A signature scheme $\SIG = (\KeyGen, \Sign_H, \Verify)$ is $\UFNMA$-secure (resp. $\sUFNMA$-secure) against quantum adversaries if it is $0-\UFCMA$-secure (resp. $0-\sUFCMA$-secure) as defined in Definition \ref{def:UF} (resp. \ref{def:sUF}).
\end{definition}

\begin{definition}[CMA-Adversary Algorithm]\label{def:CmaAdv}
    Let $X, M, Y, S$ be finite Abelian groups and $H: X \times M \rightarrow Y$ a function. 
    
    Now, let $\Sign_H: M \rightarrow S$ be a classical-query function that makes $\nu \coloneqq O(1)$ classical queries to $H$ and produces a signature $s \in S$ for message $m \in M$. That is, there exists unitaries $U_1, U_2, ... U_{2\nu}$ such that for initial state $\ket{\phi_0}$:
    \begin{align}
        \Sign_H\ket{\phi_0} \coloneqq U_{2\nu}O^HU_{2\nu-1}...U_2O^HU_1\ket{\phi_0}
    \end{align}
    where $O^H\ket{x}\ket{y} \coloneqq \ket{x}\ket{y + H(x)}$ for any tuple where $x \in X, y \in Y$. Note that $\Sign_H$ is permitted a classical query to $H$ denoted by $O^H$, not a quantum query $O^{\ket{H}}$.
    
    Next, we define a CMA-adversary algorithm $\calA^{\ket{H}, \Sign_H}$ as an algorithm that makes up to $Q$ quantum queries to $H$ and up to = $Q$ classical queries to $\Sign_H$ as a sequence of $2Q$ unitaries $A_1, A_2, ..., A_{2Q}$ and an initial state $\ket{\psi_0}$. Then, $\calA$ after $j \in [2Q]$ queries produces the state:
    \begin{align}
        \ket{\psi_j^H} \coloneqq \calA^{\ket{H}, \Sign_H}\ket{\psi_0} = A_{j}\mathcal{O}^VA_{j-1}...A_2\mathcal{O}^VA_1\mathcal{O}^V\ket{\psi_0},
    \end{align}
    where $V$ may be any of $\ket{H}, \Sign_H$ or the identity and $O^{\Sign_H}\ket{m}\ket{s}\coloneqq \ket{m}\ket{s+ \Sign_H(m)}$.
\end{definition}

Next we will build up mathematical definitions for the computational problems that underlie FALCON.

\begin{definition}[Lattice and Lattice Basis]\label{def:basis}
    A lattice $\Lambda \subseteq \R^n$ is a discrete subgroup of $\R^n$ generated by a basis. A basis $B \coloneqq [b_1|b_2|...|b_{n}]$, is any set of $n$ linearly independent column vectors such that \cite{DecadePeikert}:
    \begin{align}
        \Lambda = \{\sum_{i=1}^nx_ib_i:x_i \in \mathbb{Z}\forall i\in[n]\} = \{B\cdot x:x\in \Z^n\}
    \end{align}
\end{definition}

\begin{definition}[Gram-Schmidt Norm]\label{def:gsNorm}
    Let $\Lambda$ be a lattice with full-rank $n$ basis $M$. Let $\mathsf{MaxCol}(M)$ denote the largest Euclidean length from the columns of $M$. Then, we define the Gram-Schmidt norm of $M$ as:
    \begin{align}
        \gs{M} \coloneqq \min_{Z \in \mathcal{U}_n}\left(\mathsf{MaxCol}(MZ)\right)
    \end{align}
    where $\mathcal{U}_n$ is the set of all unital upper-triangular matrices, that is $\mathcal{U}_n \coloneqq \{Z\in \R^{n\times n}: Z \text{ upper triangular}, Z_{ii} = 1 \forall i\}$ \cite{Ducas14}.
\end{definition}

\begin{definition}[Dual Lattice]\label{def:dual}
    The dual of a lattice $\Lambda$ with basis $B$, denoted as $\Lambda^*$, is the lattice generated by the basis $B(B^TB)^{-1}$. This is equal to $(B^T)^{-1}$ if $\Lambda$ is full rank \cite{DecadePeikert}.
\end{definition}

\begin{definition}[Smoothing Parameter]\label{def:smooth}
    For any $n$-dimensional lattice $\Lambda$ and positive real $\epsilon > 0$, the smoothing parameter $\eta_\epsilon(\Lambda)$ is defined as the smallest $s$ such that \cite{DecadePeikert}:
    \begin{align}
        \sum_{x\in \Lambda^*\setminus \{0\}}e^{-\pi\abs{s x}^2} \le \epsilon
    \end{align}
\end{definition}

\begin{lemma}[{\cite[Corollary 2.8]{GPV}}]\label{lem:latticeSmooth}
Let $\Lambda$, $\Lambda'$ be $n$-dimensional lattices with $\Lambda' \subseteq \Lambda$. Then, for any $\epsilon \in (0, 1/2)$, $\gamma \ge \eta_\epsilon(\Lambda')$, and $c \in \R^n$:
\begin{align}
    \Delta(D_{\Lambda, \gamma, c} \mod \Lambda', U(\Lambda \mod \Lambda')) \le 2\epsilon
\end{align}
for $U(\Lambda \mod \Lambda')$ the uniform distribution over $\Lambda \mod \Lambda'$.
\end{lemma}

\begin{definition}[Minimum Distance]\label{minDef}
    Let $\Lambda$ be a lattice, then we define the minimum distance $\lambda_1$ of $\Lambda$ as\cite{DecadePeikert}:
    \begin{align}
        \lambda_1(\Lambda) &\coloneqq \underset{v\in \Lambda \setminus\{0\}}{\min}\left(\abs{v}\right).
    \end{align}
\end{definition}

\begin{definition}[$\gamma$-Approximate Shortest Vector Problem]\label{prob:SVP}
    Let $n \in \N$, $\gamma$ be a function of $n$, and $B$ be a basis for an $n$-dimensional lattice $\Lambda$. Then, the advantage of a quantum query algorithm $\calA$ for solving $\gamma-\SVP$ is defined as:
    \begin{align}
        \Adv^{\gamma-\SVP}(\calA) \coloneqq \Pr\Bigl[0 < \abs{v}\le\gamma(n)\lambda_1(\Lambda)\mid v \leftarrow \calA_{\gamma}(B)\Bigr].
    \end{align}
    This is simply called the Shortest Vector Problem or $\SVP$ when $\gamma(n) = 1$ \cite{DecadePeikert}.
\end{definition}

\begin{definition}[Inhomogeneous Short Integer Solution]\label{prob:ISIS}
Let $k,l,\gamma \in \mathbb{N}$. The advantage of a quantum query algorithm $\calA$ for solving $\ISIS_{k,l,\gamma}$ is defined as \cite{DecadePeikert}:
\begin{equation}
\Adv^{\ISIS}_{k,l,\gamma}(\calA) \coloneqq \Pr\Bigl[A\cdot y = t \mod q \wedge \vert y\vert\le \gamma\mid A\sample \Z_q^{k\times l}, t\sample \Z_q^k, \ y \leftarrow\calA(A, t) \Bigr].
\end{equation}
\end{definition}

\begin{definition}[Short Integer Solution]\label{prob:SIS}
Let $k,l,\gamma \in \mathbb{N}$. The advantage of a quantum query algorithm $\calA$ for solving $\SIS_{k,l,\gamma}$ is defined as \cite{DecadePeikert}:
\begin{equation}
\Adv^{\SIS}_{k,l,\gamma}(\calA) \coloneqq \Pr\Bigl[A\cdot y = 0 \mod q \wedge 0 < \abs{y}\le \gamma\mid A\sample \Z_q^{k\times l}, \ y \leftarrow\calA(A) \Bigr].
\end{equation}\end{definition}

The $\NTRU$ learning problem (defined below) asks whether an efficient algorithm can distinguish a lattice constructed using an $\NTRU$ public key from a comparable uniformly random lattice. It is therefore useful to formalize $\FALCON$'s security assumptions as the hardness of this distinguishing task together with the hardness of the more conventional lattice problems (i.e., $\LWE$ and $\SIS$) rather than as assumptions about the hardness of $\LWE$ or $\SIS$ over $\NTRU$ lattices, which would be non-standard.

\begin{definition}[NTRU Learning]\label{prob:ntru}
Let $\sigma_{f, g} \in \mathbb{R}$, then, the advantage of a quantum query algorithm $\calA$ for solving $\NTRU_{\sigma_{f, g}}$ is defined as \cite{DecadePeikert}:
\begin{align}\label{eq:ntru}
    \begin{split}
        \Adv^{\NTRU}_{\sigma_{f, g}}(\calA) \coloneqq \big|\Pr[b=1 \mid f, g \sample D_{\Z^n, \sigma_{f, g}, 0}, h\coloneqq gf^{-1}\mod q, b\leftarrow\calA(h)]& \\
        \qquad- \Pr[b=1\mid h\sample R_{q, n}, b\leftarrow \calA(h)]&\big|.
    \end{split}
\end{align}
\end{definition}

\textbf{Hermite Lattice Problems:} The above lattice problems also admit Hermite-form variants wherein the operators take the form $[I|A]$ for $A \in \Z_q^{k\times m}$ and $k, m \in \N$ \cite{DecadePeikert}. Note that the probability that a uniformly random $A \sample \Z_q^{k \times (k+l)}$ is of rank $k$ --- and thus has a row-echelon form $[I|B']$ for some $B'$ --- is lower bounded by $1 - O(q^{-k-1})$. Then, since the time complexity of row reducing $A$ is $O(n\log(q)\cdot k l \min(k, l))$, it follows that the existence of an efficient algorithm solving the Hermite form of a lattice problem implies the existence of an efficient algorithm solving the non-Hermite form, and vice-versa \cite{DilithiumQROM}.

\textbf{NTRU Lattice Problems:} The above lattice problems can also be defined over NTRU lattices with operators taking the form $\begin{bmatrix}1 &h \\ 0 &q\end{bmatrix}$ for $h \coloneqq f^{-1}g\mod q$ and $fG - gF = q \mod \phi$ where $f, g, F, G$ are small polynomials over $R_{q, n}$ \cite{DecadePeikert}. Since non-$\NTRU$ lattice problems generalize their $\NTRU$-counterparts, any efficient algorithm solving the former can be used to construct an efficient algorithm for solving the latter. In this sense, $\NTRU$ variant problems are no harder than their non-$\NTRU$ counterparts.

\textbf{Ring Lattice Problems:} The above lattice problems can also be applied to operators, target vectors, and solutions with elements in $R_{q, n}$ rather than $\Z_q$. As unstructured lattice problems generalize their ring-lattice counterparts, any efficient algorithm solving the former can be used to construct an efficient algorithm for solving the latter. In this sense, ring-lattice problems are no harder than their unstructured-lattice counterparts. However, as it is generally difficult to find an advantage from the ring structure of $R_{q, n}$, a common heuristic is to assume that ring problems are exactly as hard as their unstructured equivalents \cite{DecadePeikert}.

The following definition is a type of interactive protocol that will arise in intermediate steps in the proof of our main result.

\begin{definition}[Challenge-Response Protocol]\label{def:intPro}
    A Challenge-Response Protocol is a $4$-tuple $(M, C, R, V)$ such that $M, C, R$ are alphabets and $V: M\times C \times R \rightarrow \{0, 1\}$ is a function. 
\end{definition}
A challenge-response protocol $(M, C, R, V)$ describes an interaction between two parties, a Prover $\mathcal{P}$ and a Verifier $\mathcal{V}$, as follows:
\begin{enumerate}
    \item Prover uses an algorithm $\mathcal{P}_1$ to generate a message $m \in M$, and sends it to Verifier.

    \item Verifier generates a uniformly random element $c \leftarrow C$ and sends it to Prover.

    \item Prover uses a second algorithm $\mathcal{P}_2$ to generate a response $r \in R$ and sends it to Verifier.

    \item Verifier computes the function $V(m, c, r)$.
\end{enumerate}
The tuple $(m,c,r)$ is referred to as the transcript of the interaction. If $V(m,c,r) = 1$, we refer to the transcript as an accepting transcript. If not, we refer to the transcript as a rejecting transcript.

Given a challenge-response protocol, we may consider a prover algorithm seeking to produce an accepting transcript.

\begin{definition}[Adversary Algorithm]\label{def:SigAdv}
    Let $X$ be an alphabet and $Q \in \N$. Then, for any challenge-response protocol $\Phi_0 \coloneqq (M, C, R, V_H)$ where $H: X \rightarrow M$ is an oracle accessed by $V_H$, we define a $\Phi_0$-adversary algorithm $\calS^{\ket{H}}_{Q} \coloneqq (S_1, S_2)$ as an algorithm of the following form:
    \begin{enumerate}
        \item $i \sample [Q]$
        \item $\ket{x}, T \leftarrow S_1^{\ket{H}}$ where $\ket{x} \in \mathbb{C}^X$ and $T$ is an arbitrary work register state
        \item $x_0 \coloneqq \mathbf{M}\ket{x}$ for $\mathbf{M}$ a destructive measurement outputting $x_0 \in X$
        \item $\Theta \sample C$
        \item $\ket{y} \leftarrow S_2^{\ket{H[\Theta]}}(x_0, \Theta, T)$
    \end{enumerate}
    Above, $S_1$ makes $i$ quantum queries to $H$ while $S_2$ makes $Q-i$ quantum queries to $H[\Theta]$. Here, $H[\Theta](x) \coloneqq \begin{cases}H(x) &x\ne x_0 \\ \Theta &x=x_0\end{cases}$.
\end{definition}

The relevance of this definition to adaptive reprogramming will become clear in Section \ref{sec:tightness}, where we use this framework to analyze the effect of reprogramming on such protocols.
\section{Tightness of Reprogramming Reductions for Hash-and-Sign Signatures}\label{sec:tightness}

One of the techniques that we will utilize in our security proof is the Measure \& Reprogram algorithm from Don et al.~\cite{Reprogram}. This technique utilizes a reprogrammed oracle. In general, this result, and others described as 'adaptive reprogramming' attempt to bound how much information or how much advantage can be gained by allowing quantum queries to an oracle instead of classical ones. 

This is where, given a random oracle $H: X\rightarrow Y$ for alphabets $X, Y$ and values $x_0 \in X, \Theta\sample Y$:
\begin{align}
    H[\Theta](x) \coloneqq \begin{cases}
        H(x) &x\ne x_0 \\
        \Theta &x = x_0
    \end{cases}.
\end{align}

Thus, their algorithm provides the following lemma:
\begin{lemma}[Measure \& Reprogram~{\cite[Adapted from Theorem 2]{Reprogram}}]\label{lem:measure_and_reprogram}Let $\calA$ be a quantum query algorithm making $Q$ queries to a uniformly random $H: X\rightarrow Y$ before outputting some $x \in X$ and $z \in Z$ for $X, Y, Z$ alphabets. Let $V_H: X \times Y\times Z\rightarrow \{0, 1\}$ be a verifier function. For any such $\calA$ and $V_H$ we define a challenge-response protocol $\Phi_0 \coloneqq (X, Y, Z, V_H)$. Then, there exists a $\Phi_0$-adversary algorithm $\calS_{Q}^{\ket{H}} \coloneqq (S_1, S_2)$ (Definition \ref{def:SigAdv}) such that:
\begin{align}
\begin{split}
    &\Pr_{\Theta \in Y}[x=x_0\wedge V(x, \Theta, z)\mid x\leftarrow S_1, z\leftarrow S_2(x, \Theta)] \\
    &\qquad\qquad\ge \frac{1}{(2Q+1)^2}\Pr_{H\in \Func(X,Y)}[x=x_0\wedge V(x, H(x), z):(x, z)\leftarrow \calA^{\ket{H}}],
\end{split}
\end{align}
for arbitrary $x_0 \in X$. Equivalently, in terms of the success probability of $\calS$ and $\calA$ for satisfying the verifier $V_H$ we may write:
\begin{align}
    \sigma_\calS \ge \frac{1}{(2Q+1)^2}\sigma_\calA.
\end{align}
Further, $\Time(\calS)$ is polynomial with respect to $Q, \log(\abs{X})$, and $\log(\abs{Y})$.
\end{lemma}

It has previously been demonstrated that the above result is asymptotically optimal for Fiat-Shamir signatures \cite{Reprogram}, but we now demonstrate that this is optimal for Hash-and-Sign signature schemes as well. To that end, we define two constructions which are derived from Hash-And-Sign signature constructions. For a hash function $H:M \rightarrow C$, we construct a Hash-And-Sign signature scheme by constructing signatures of the form $(m, r)$ such that $H(m), r$ satisfies some verifying function $V: C\times R \rightarrow \{0, 1\}$. 

\begin{definition}[Predicate Problem]
    Let $M, C, R$ be alphabets. Then, given a verifier $V: C\times R \rightarrow \{0, 1\}$ and hash function $H: M \rightarrow C$, the predicate problem is the task of finding $(m, r) \in M \times R$ such that $V(H(m), r) = 1$.
\end{definition}

\begin{definition}[Subset Search Algorithms]
    A $Q$-quantum-query quantum algorithm $\calA_Q^{\ket{\mathcal{H}}}$ is a $(\epsilon, Q, \sigma$)-subset search algorithm if and only if for $M, C, C'$ finite, non-empty sets where $C' \subset C$ and $\abs{C'}/\abs{C} \le \epsilon$:
    \begin{align}
        \Pr[m \in M \wedge H(m) \in C'\mid m \leftarrow \calA_Q^{\ket{H}}] \ge \sigma.
    \end{align}
    Here, $H \sample \Func(M, C)$.
\end{definition}

The predicate problem can be understood as another way of expressing the problem of forging a valid message/signature pair $(m, r)$ that satisfies the verifier of a Hash-and-Sign signature scheme. We then recognize that a subset search algorithm formalizes any algorithm seeking to solve the predicate problem by seeking $m$ such that $H(m)$ lies in the subset $C' \coloneqq \{c \in C \mid V(c, r)=1\}$ for a randomly selected $r$. Therefore, when we show that subset search algorithms cannot have a reprogramming result with better than a $O(Q^{-2})$ loss in success probability, it follows that in general the adversaries of Hash-and-Sign schemes cannot have better reprogramming results.

Using these definitions, we claim:
\begin{lemma}\label{lem:tight}
    Given a $Q$-quantum query algorithm $\calA^{\ket{H}}$ that solves the subset search algorithm, there exists no $\Phi_0$-adversary algorithm $\calS_{Q}^{H}$ for $\Phi_0 \coloneqq (M, C, R, V)$ as given above such that:
    \begin{align}
        \sigma_\calS = \Theta(Q^{-(2-\delta)})\sigma_\calA
    \end{align}
    for any $0 < \delta \le 1$
\end{lemma}

\begin{proof}
We will prove this by assuming the contradiction. First, assume that there exists an improved Measure \& Reprogram algorithm wherein every $Q$-quantum query algorithm $\calA^{\ket{H}}$, implies the existence of a $\Phi_0$-adversary algorithm $\calS_{Q}^{H}$ for a challenge-response protocol $\Phi_0 \coloneqq (M, C, R, V)$ with success probability at satisfying the verifier $V$ with $m_0 \in M$:
\begin{align}
\begin{split}
    &\Pr_{\Theta \in C}[m=m_0\wedge V(m, \Theta, r)\mid m\leftarrow S_1, r\leftarrow S_2(m_0, \Theta)] \\
    &\qquad\qquad= \Theta(Q^{-(2-\delta)})\Pr_{H\in \Func(M,C)}[m=m_0\wedge V(m, H[\Theta](m), r):(m, r)\leftarrow \calA^{\ket{H}}].
\end{split}
\end{align}
for some $0 < \delta \le 1$ where $\delta=0$ gives the existing Measure \& Reprogram result while $\delta=1$ matches the best existing results from classical rewinding. We may use this assumption to imply a classical search algorithm more efficient than existing results.

Now, let $\calA^{\ket{H}}$ be a Grover's search algorithm that returns some $m \in M$ such that $H(m) \in C'$. The multi-target variation of this algorithm (that is, amplitude amplification \cite{AmplitudeSearch}) produces an $(\epsilon, O(1/\sqrt{\epsilon}), 1)$-subset search algorithm for $\abs{C'}/\abs{C}\le\epsilon$.

Then, using the improved Measure \& Reprogram result that we assumed above, such a subset search algorithm implies the existence of an algorithm $\calS_{Q}^{H}$ such that:
\begin{align}
    \sigma_\calS = \Theta(Q^{-(2-\delta)})\sigma_\calA < O(\epsilon),
\end{align}
as by definition, $0 < \epsilon < 1$. Now, 

Thus, we find a search algorithm that given a uniformly random $\Theta$ from the set $C$ finds an element $m \in M \ni H(m) \in C'$ with an asymptotic success probability greater than $\epsilon$. As this is clearly impossible, we conclude that no Measure \& Reprogram algorithm for subset search problems (and by construction Hash-and-Sign signatures) improved in this fashion may exist, as claimed.
\qed\vspace{\baselineskip}\end{proof}

This gives a simpler proof that a universal improvement to Measure and Reprogram's result like $O(Q^{-(2-\delta)})$, for any fixed \(\delta>0\), would contradict known classical search bounds. This also provides a new proof that it is specifically optimal when applied to the cryptanalysis of Hash-And-Sign signature schemes.
\section{QROM security proof for FALCON}
\subsection{Security Games}\label{sec:Games}
Now, we consider the QROM security of the $\FALCON^+$ scheme. To this end, we will compare a series of security games. Note that, in FALCON, the parameters are chosen to optimize signing efficiency, subject to constraints on the Gram-Schmidt norm of the NTRU basis. In particular, the parameters are selected such that
\begin{align}
\gs{B} \leq 1.1658\sqrt{q},
\end{align}
where the constant (1.1658) is the optimized bound used in the FALCON parameter selection. We use this bound as needed throughout the following analysis. We also define:
\begin{align}\label{eq:pGS}
    p_{GS} \coloneqq \Pr[\max\left(\abs{(g, -f)}, \abs{\left(\frac{q f^*}{ff^* + gg^*}, \frac{qg^*}{ff^* + gg^*}\right)}\right)\le 1.1658\sqrt{q}: f, g\leftarrow D_{R_{q, n}, \sigma_{f, g}, 0} | f^{-1} \in R_{q, n}].
\end{align}
According to the authors of $\FALCON$, for $\gs{B} \le 1.1658\sqrt{q}$, we may approximate $p_{GS} \approx 1/4$ \cite{FALCON}.

\begin{theorem}
Let $\calA$ be a $Q$-quantum query algorithm. Then for any positive $(\beta, \sigma_{f, g}, \sigma) \in \mathbb{R}^3$, $\lambda \in \N, \zeta \coloneqq 2^{-\lambda}/4n$, satisfying $\sigma \ge \max\left(1.1658\sqrt{q}\eta_\zeta(\Z), 1.1658\sqrt{q}\sqrt{\ln(2n(1+1/\delta))/\pi}, \eta_{\epsilon/n}(\Z)\right)$ there exists $\ISAlg, \calN$ efficient quantum algorithms given access to $\calA$, such that:
\begin{align}\label{eq:theo}
    \Adv^{Q-\UFCMA}_{\FALCON^+}(\calA) \le (4Q+1)^2p_{GS}^{-1}\left(\Adv_{\sigma_{f, g}}^{\NTRU}(\calN) + \Adv_{1, \beta}^{\RISIS}(\ISAlg)\right) + Q\left(\frac{\chi + 2^{-\lambda}}{2-2\chi} + (4Q+1)^2(4\delta + 2^{-\lambda})\right)
\end{align}
Here, $\delta\in(0, 1/2), \epsilon \in (0, 1/4)$, and $\chi\coloneqq 2^{-\lambda} + \frac{1+\epsilon}{1-\epsilon}2^{-2n}$.
\end{theorem}
We establish this result through a series of lemmas.

\paragraph{Game $\Gufcma$:} Game $\Gufcma$ is the unforgeability condition of $\FALCON^+ = (\KeyGen, \Sign^+_H, \Verify)$ as defined in Figure \ref{fig:falcon_simplified} under chosen message attacks as defined in Def'n \ref{def:CMA}.
Thus, for a $Q$-query CMA Adversary Algorithm $\UFAlg$ as given in Def'n \ref{def:CmaAdv}:
\begin{align}
\begin{split}
    \Pr[\Gufcma^\UFAlg] &\coloneqq \Adv^{Q-\UFCMA}_{\FALCON^+}(\UFAlg), \\
    &= \Pr_{pk, sk \leftarrow \KeyGen}[\Verify(pk, m, \sigma), \ m \notin \textup{SignQ} \mid (m, \sigma)\leftarrow\UFAlg^{\ket{H}, \Sign^+_H(sk, \cdot)}(pk)].\label{eq:Gufcma}
\end{split}
\end{align}

\paragraph{Game $\Gsignrep$:} Game $\Gsignrep$ is the same as $\Gufcma$ except we replace the oracle $\Sign^+_H$ with an oracle $\Sign_H$ motivated below. 
\begin{align}
\begin{split}
    \Pr[\Gsignrep^\SRAlg] &\coloneqq \Pr_{pk, sk \leftarrow \KeyGen}[\Verify(pk, m, \sigma), \ m \notin \textup{SignQ} \mid (m, \sigma)\leftarrow\SRAlg^{\ket{H}, \Sign_H(sk, \cdot)}(pk)].\label{eq:Gsignrep}
\end{split}
\end{align}

In $\Sign^+_H$ as observed in Figure \ref{fig:falcon_simplified}, some constant number of repetitions are allowed to ensure that the pre-sampling algorithm $\Klein$ returns a sufficiently small $s_1, s_2$. Alternatively, in $\Sign_H$, also from Figure \ref{fig:falcon_simplified}, we select a single uniformly random $r$ then allow a constant number of repetitions to $\Klein$ with different random tapes until we receive a sufficiently small $s_1, s_2$. In both cases, information about unsuccessful $\Klein$ sampling attempts is not revealed via the signing oracles. Therefore, distinguishing between our first two games would require an algorithm making $Q$ queries to these distinct distributions to be able to distinguish them. To that end, we first discuss a few lemmas regarding the $\Klein$ sampler used by both $\Sign^+_H$ and $\Sign_H$. First, from Ducas:

\begin{lemma}[Klein Sampler Indistinguishability]
Let $n \in \N$, $\epsilon \in (0, 1/4)$, $B \in \Z^{2n\times 2n}$, $\sigma \ge \eta_\epsilon(\Z^{2n})\gs{B}$ where $\eta_\epsilon$ denotes the smoothing parameter, and $t \in R_{q, n}^2$. Then, the $a$-Rényi divergence $\mathcal{R}_a$ between a discrete Gaussian over the lattice $\Lambda(B)$ generated by $B$ compared to the output of $\Klein(B, \sigma, t)$ satisfies:
\begin{align}
\mathcal{R}_a\left(\Klein(B, \sigma, t), D_{\Lambda(B), \sigma, t}\right) \le \left(1 + \frac{a(a-1)\delta^2}{2(1-\delta)^{a+1}}\right)^{\frac{1}{a+1}}
\end{align}
where $\delta \coloneqq \left(\frac{1+\epsilon/n}{1-\epsilon/n}\right)^n - 1 \approx 2\epsilon$ \cite{Prest}. Further \cite{Ducas}:
\begin{align}
    \Delta\left(\Klein(B, \sigma, t), D_{\Lambda(B), \sigma, t}\right) \le 2^{-\lambda},\label{eq:ks-sd}
\end{align}
provided that for $\lambda \in \N, \zeta \coloneqq 2^{-\lambda}/4n$, $\sigma \ge \eta_\zeta(\Z) \gs{B}$ \cite{Ducas}. 
\end{lemma}

The authors of $\FALCON$ and $\FALCON^+$ utilize the former result, while our efforts engage the latter on the basis that statistical distance is more conventional and broadly intuitive \cite{FALCON, CloserLook}.

\begin{lemma}\label{lem:klein}
Let $k \in \N$, $B \in R_{q, n}^{k \times k}$ of rank $r$, $t, t' \in R_{q, n}^k$. Then:
\begin{align}
    \abs{\Pr[|v|^2 > \beta^2|v\leftarrow \Klein(B, \sigma, t)] - \Pr[|v|^2 >\beta^2|v\leftarrow \Klein(B, \sigma, t')]} \le 2^{1-\lambda} + \frac{1+\epsilon}{1-\epsilon}2^{-rn}
\end{align}
for $\beta \coloneqq \sigma \sqrt{rn}$ provided that $\sigma \ge \max(\eta_\zeta(\Z)\gs{B}, \eta_{\epsilon/n}(\Z))$ where $\Lambda$ is the lattice generated by $B$ and $\lambda \in \N$ where $\zeta \coloneqq 2^{-\lambda}/4n$
\end{lemma}

\begin{proof}
    First, from Eq'n \ref{eq:ks-sd} and considering a single term:
    \begin{align}
        \Pr[|v|^2>\beta^2|v\leftarrow \Klein(B, \sigma, t)] \in [\Pr[|x-t|^2>\beta^2|x\leftarrow D_{\Lambda(B), \sigma, t}] - 2^{-\lambda},  \Pr[|x-t|^2>\beta^2|x\leftarrow D_{\Lambda(B), \sigma, t}] + 2^{-\lambda}]
    \end{align}
    for $\lambda \in \N$ where $\zeta \coloneqq 2^{-\lambda}/4n$, and $\sigma \ge \eta_\zeta(\Z)\gs{B}$ and $\Lambda(B)$ the lattice generated by the column vectors of $B$ as described in Def'n \ref{def:basis}. Now, from Lemma 4.4, Micciancio and Regev state for $\beta \coloneqq \sigma\sqrt{r n}$ where $B$ is rank $r$: \cite{1366257}:
    \begin{align}
        \Pr[|x-t|^2>\beta^2|x\leftarrow D_{\Lambda(B), \sigma, t}] \le \frac{1+\epsilon}{1-\epsilon}2^{-rn}
    \end{align}
    provided that $\sigma \ge \eta_\epsilon(\Lambda(B))$. Recall $\eta_\epsilon(\Lambda(B) \le \eta_{\epsilon/n}(\Z)$ as discussed in the proof of Lemma \ref{lem:klein}. As this is independent of the target vector $t$, it follows that:
    \begin{align}
        \sup_{t, t' \in R_{q, n}^{k}}|\Pr[|x-t|^2>\beta^2|x\leftarrow D_{\Lambda(B), \sigma, t}] - \Pr[|x-t|^2>\beta^2|x\leftarrow D_{\Lambda(B), \sigma, t'}]| \le \frac{1+\epsilon}{1-\epsilon}2^{-rn}
    \end{align}
    Thus, the claim follows via the triangle inequality for all $t, t'$ taking the substitution $x=v+t$. Observe that $\Lambda(B) = \Lambda([1|h_B]) = \Z^n$ and $\eta_\epsilon(\Z^n) \le \eta_{\epsilon/n}(\Z)$ \cite{DecadePeikert}.
\end{proof}

Now, we can set out to prove using the lemma above that access to either $\Sign^+_H$ or $\Sign_H$ is statistically indistinguishable for our purposes. Formally:

\begin{lemma}
    Let $\UFAlg^{\ket{H}, \Sign^+_H}$ be any $Q$-quantum CMA-Adversary algorithm. Here, $\Sign^+_H$ and $\Sign_H$ are described in Figure \ref{fig:falcon_simplified}. Then, for $\Pr[\Gufcma^\UFAlg]$ as defined in Equation \ref{eq:Gufcma}, there exists $\SRAlg$ such that for $\Pr[\Gsignrep^\SRAlg]$ as defined in Equation \ref{eq:Gsignrep}
    \begin{align}
        \abs{\Pr[\Gsignrep^\SRAlg] - \Pr[\Gufcma^\UFAlg]} \le \frac{Q}{2}\frac{\chi + 2^{-\lambda}}{1-\chi}
    \end{align}
    Here, $\chi \coloneqq 2^{-\lambda} + \frac{1+\epsilon}{1-\epsilon}2^{-2n}$, $\lambda \in \N$, $\zeta \coloneqq 2^{-\lambda}/4n$, $\beta \coloneqq \sigma \sqrt{rn}$, $\epsilon \in (0, 1/4)$ and provided that $\sigma \ge \max(\eta_\zeta(\Z)1.1658\sqrt{q}, \eta_{\epsilon/n}(\Z))$.
\end{lemma}

\begin{proof}
    Given an algorithm $\UFAlg$, we may construct an efficient algorithm $\SRAlg$ by running $\UFAlg$ and forwarding its signing queries to the $\Sign_H$ oracle. Algorithms $\UFAlg^{\ket{H}, \Sign^+_H}$ and $\SRAlg^{\ket{H}, \Sign_H}$ therefore both make no more than $Q$ black-box queries to their respective signing oracles. Let $\mathrm{Accept}$ denote the event that $|S_1, S_2|^2 \le\beta^2$. Then, let $K_r$ denote the distribution of $S_2$ induced by $\Klein(B,\sigma,r)$, and let $K_r^{\text{accept}}$ denote the corresponding subdistribution restricted to accepted outputs for any $r \in \{0, 1\}^k$. That is:
    \begin{align}
        K_r^\text{accept}(s_2) &\coloneqq \Pr[s_2=S_2 \wedge \textrm{Accept}| (S_1, S_2)\leftarrow \Klein(B, \sigma, r)] \\ 
        K_r(s_2) &\coloneqq \Pr[s_2=S_2 | (S_1, S_2)\leftarrow \Klein(B, \sigma, r)]
    \end{align}
    Then, given any query to $\Sign_H$, $\Sign_H$ outputs a sample $(r, s_2)$ from the distribution:
    \begin{align}
        P(r, s_2) &\coloneqq \Pr_{R \in \{0, 1\}^k}[R=r]\Pr[S_2=s_2|S_2\leftarrow K_r, \text{Accept}] \\
        &= 2^{-k}\frac{K_r^{\text{accept}}(s_2)}{\Pr_{S_2\leftarrow K_r}[\mathrm{Accept}]}
    \end{align}
    $\Sign_H$ selected a value $R$ uniformly at random, then for that fixed $r$, generates repeated random $s_2$ values until one is sufficiently small. Given any query to $\Sign^+_H$, $\Sign^+_H$ outputs a sample $(r, s_2)$ from the distribution:
    \begin{align}
        X(r, s_2) &\coloneqq \Pr_{R \in \{0, 1\}^k}[(R, S_2)=(r, s_2)|R\leftarrow \{0, 1\}^k, S_2\leftarrow K_R, \text{Accept}] \\
        &= 2^{-k}\frac{K_r^{\text{accept}}(s_2)}{\mathbb{E}_{r\in \{0, 1\}^k}[\Pr_{S_2\leftarrow K_r}[\mathrm{Accept}]]}
    \end{align}
    Unlike before, $\Sign^+_H$ selects a uniformly random $R, S_2$ then keeps or discards both values depending on the norm of $S_2$. As a result, this distribution is relatively biased towards $R$ values which cause $\Klein$ to output small $S_2$ values with higher probability than other $R$ values. However, we will now demonstrate that these two output distributions are statistically indistinguishable. To begin:
    \begin{align}
        \Delta(P, X) &\coloneqq \frac{1}{2}\sum_{r, s_2}\abs{P(r, s_2)-X(r, s_2)} \\
        &= \frac{1}{2}\sum_{r}2^{-k}\abs{\frac{1}{\Pr_{S_2\leftarrow K_r}[\mathrm{Accept}]} - \frac{1}{\mathbb{E}_{r\in \{0, 1\}^k}[\Pr_{S_2\leftarrow K_r}[\mathrm{Accept}]]}}\sum_{s_2}K_r^{\text{accept}}(s_2)
    \end{align}
    Next, by our definition for $K_r^{\text{accept}}$:
    \begin{align}
        \Delta(P, X) &= \frac{1}{2}\sum_{r}2^{-k}\abs{1 - \frac{\Pr_{S_2\leftarrow K_r}[\mathrm{Accept}]}{\mathbb{E}_{r\in \{0, 1\}^k}[\Pr_{S_2\leftarrow K_r}[\mathrm{Accept}]]}}
    \end{align}
    And if we let $k_r \coloneqq \Pr_{S_2\leftarrow K_r}[\mathrm{Accept}]$ and $\bar{k}$ be the expectation value of $k_r$ over $r\in \{0, 1\}^k$:
    \begin{align}
        \Delta(P, X) &= \frac{\mathbb{E}_{r\in \{0, 1\}^k}[\abs{k_r - \bar{k}}]}{2\bar{k}}.
    \end{align}
    Note that the above expression requires $k_r > 0 \forall r$. Using a similar method demonstrated in \ref{lem:klein} we quickly observe that:
    \begin{align}\label{eq:lowK}
        k_r \ge 1-2^{-\lambda}-\frac{1+\epsilon}{1-\epsilon}2^{-2n} > 0.
    \end{align}
    It also follows that $\bar{k} >0$. Now, recall that Lemma \ref{lem:klein} implies:
    \begin{align}
        \abs{k_r - \bar{k}} \le 2^{1-\lambda} + \frac{1+\epsilon}{1-\epsilon}2^{-rn}
    \end{align}
    as acceptance and rejection are complementary events. Thus, we conclude:
    \begin{align}
        \Delta(P, X) \le \frac{1}{2\bar{k}}\left(2^{1-\lambda} + \frac{1+\epsilon}{1-\epsilon}2^{-rn}\right).
    \end{align}
    Therefore, using a hybrid argument and accounting for $Q$ queries by Bernoulli's inequality we find that:
    \begin{align}
        \abs{\Pr[\Gsignrep^\SRAlg] - \Pr[\Gufcma^\UFAlg]} \le \frac{Q}{2\bar{k}}\left(2^{1-\lambda} + \frac{1+\epsilon}{1-\epsilon}2^{-2n}\right).
    \end{align}
    Recall that $B$ is of full rank $2$ and $\gs{B} \le 1.1658\sqrt{q}$ for efficiency. Finally, utilizing Eq'n \ref{eq:lowK}, and letting $\chi \coloneqq 2^{-\lambda} + \frac{1+\epsilon}{1-\epsilon}2^{-2n}$:
    \begin{align}
        \abs{\Pr[\Gsignrep^\SRAlg] - \Pr[\Gufcma^\UFAlg]} \le \frac{Q}{2}\frac{\chi + 2^{-\lambda}}{1-\chi}
    \end{align}
\qed\vspace{\baselineskip}\end{proof}

\paragraph{Game $\Gmeasrep$:} Let $\Theta\sample R_{q, n}$. Then, Game $\Gmeasrep$ is the success probability of a quantum algorithm $\MRAlg^{\ket{H[\Theta]}, \Sign_{H[\Theta]}}$ satisfying $\Verify$ given access to the public key $pk$, quantum-query access to $H[\Theta]$ reprogrammed during the $i$th query for $i\sample [Q]$, as well as black-box access to the unitaries $B_1, B_2$ composing $\Sign_{H[\Theta]}$ (as described above):
\begin{align}
    \Gmeasrep \coloneqq \Pr_{pk, sk \leftarrow \KeyGen}[\Verify_{H[\Theta]}(pk, m, \sigma), \ m \notin \textup{SignQ} \mid (m, \sigma)\leftarrow\MRAlg_Q^{\ket{H[\Theta]}, B_1, B_2}(pk)].\label{eq:Gmeasrep}
\end{align}

Our first lemma will apply the Measure and Reprogram algorithm to the CMA-Adversary Algorithm in Game $\Gsignrep$ in order to relate $\Gsignrep$ to $\Gmeasrep$.
\begin{lemma}
    Let $\SRAlg^{\ket{H}, \Sign_H}$ be any $Q$-quantum CMA-Adversary algorithm. Then, for $\Pr[\Gsignrep^\SRAlg]$ as defined in Equation \ref{eq:Gsignrep} let $\Phi_\mathcal{O}$ be a challenge-response protocol $(\{0, 1\}^k, R_{q, n}, R_{q, n}, \Verify_{\mathcal{O}})$. Then for $\Pr[\Gmeasrep^\MRAlg]$ as given Equation \ref{eq:Gmeasrep}, there exists a $\Phi_{H[\Theta]}$-adversary algorithm $\MRAlg^{\ket{H}, B_1, B_2}_Q$ with access to $B_1, B_2$ as black-boxes such that:
    \begin{align}
        \Pr[\Gmeasrep^\MRAlg] \ge \frac{1}{(4Q+1)^2}\Pr[\Gsignrep^\SRAlg].
    \end{align}
\end{lemma}

\begin{proof}
Given a CMA-Adversary Algorithm $\SRAlg^{\ket{H}, \Sign_H}$ as given in Definition \ref{def:CmaAdv} with $\nu$ the number of queries to $H$ made by $\Sign_H$ fixed at $1$. WLOG, we may write:
\begin{align}
    \ket{\psi_j^H} \coloneqq \SRAlg^H\ket{\psi_0} = A_jB_2O^{H}B_1A_{j-1}O^{\ket{H}}...A_2B_2O^{H}B_1A_1O^{\ket{H}}\ket{\psi_0},
\end{align}
where $B_1, B_2$ act on the appropriate sub-register. Then, we define a new set of unitaries $R_i \coloneqq B_1A_i$ for odd $i \in [2Q]$ and $R_i \coloneqq B_2 A_i$ for even $i \in [2Q]$. Thus,
\begin{align}
    \ket{\psi_j^H} \coloneqq \SRAlg^{\ket{H}, \Sign_H}\ket{\psi_0} = R_jO^{\ket{H}}R_{j-1}O^{\ket{H}}...R_2O^{\ket{H}}R_1O^{\ket{H}}\ket{\psi_0},
\end{align}
We recognize this as a quantum-query algorithm. This allows us to apply the measure-and-reprogram algorithm, with a total number of queries equal to $2Q$ for $Q$ the number of queries allowed to our initial CMA-Adversary algorithm. Then, by the construction of $\Gtrapd$ and $\Gsignrep$, this proves the lemma.
\qed\vspace{\baselineskip}\end{proof}

At this point in our efforts, our algorithm $\SRAlg$ can be written as acting on some initial state $\phi_0$ with reprogramming at the $i$th query as follows:
\begin{align}
    \SRAlg^{H[\Theta], B_1, B_2}\ket{\phi_0} = R_{2Q} O^{\ket{H[\Theta]}}R_{2Q-1}...R_{i+2} O^{\ket{H[\Theta]}}R_{i+1} O^{\Theta}XR_{i}O^{\ket{H}}R_{i-1} ...R_1 O^{\ket{H}}\ket{\phi_0}
\end{align}
where $O^\Theta$, the query made immediately after a complete measurement of the query register returns some $x_0 \in X$, has a $0.5$ chance of responding with $H(x_0)$ and a $0.5$ chance of responding with $H[\Theta](x_0)$, the latter of which is - by construction - equal to $\Theta$.

Next, we will replace $H$ with a trapdoor function $H'$. Let $\gamma \in \R$, then the function $H'_{F}$ is chosen from a distribution over the set of functions $\mathsf{F}_{h_B}$ defined as follows:
\begin{enumerate}
    \item For any $f_1, f_2 \in \Func(\{0, 1\}^k, R_{q, n})^2$
    \item Let $F_1$ (resp. $F_2$) be a Box-Muller transform of $f_1$ (resp. $f_2$) such that $F_1$ (resp. $F_2$) inputs $\{0, 1\}^k$ and outputs a Gaussian $\rho_{0, \gamma}(x)$ distribution over $R_{q, n}$.
    \item Let $F(r) \coloneqq ( F_1(r), F_2(r))$
    \item $H_{F}(r) \coloneqq [1|h_B]\cdot F(r) \mod q$
    \item $\mathsf{F}_{h_B} \coloneqq \{H_{F} \mid f_1, f_2 \sample \Func(\{0, 1\}^k, R_{q, n})^2\}$
\end{enumerate}

\paragraph{Game $\Gtrapd$} This replacement produces the experiment:
\begin{align}
    \Gtrapd \coloneqq \Pr_{pk, sk \leftarrow \KeyGen}[\Verify(pk, m, \sigma), \ m \notin \textup{SignQ} \mid (m, \sigma)\leftarrow\TDAlg_Q^{\ket{H'_{F}}, B_1, B_2}(pk)].\label{eq:Gtrapd}
\end{align}
for $\TDAlg_Q^{\ket{H'_{F}}, B_1, B_2}$ a $\Phi_{H'[\Theta]}$-adversary algorithm with black-box access to $B_1, B_2$.

Before proceeding, we'll introduce some notation and properties. To that end, for $a \in R_{q, n}$ and $A \in R_{q, n}^{k\times m}$ for any positive integers $k, m$ parsed such that $A = (a_{ij})_{i, j}$ we say that $\mathsf{Rot}$ denotes:
\begin{align}
    \mathsf{Rot}(a)&\coloneqq [\tau(a)|...|\tau(X^{n-1}a)] \in \R^{n\times n}, \\
    \mathsf{Rot}(A) &\coloneqq (\mathsf{Rot}(a_{ij}))_{i,j} \in \R^{kn\times mn}.
\end{align}
In addition, for $a \in R_{q, n}^2$, we define the following notation:
\begin{align}
    \Lambda(a) &\coloneqq \{(t_1, t_2) \in (\R^{n})^2: \exists s \in R_{q, n} \ni t_i\mod q = \Rot(a_i)^T\tau(s) \forall i \in [2]\},
\end{align}
where $\tau: R_{q, n} \mapsto \R^n$ maps a polynomial in $R_{q, n}$ of the form $\sum_{i=0}^{n-1}p_iX^i$ to the corresponding coefficient vector $(p_0, p_1, ..., p_{n-1})$ in $\R^n$.

We'll also define some of the notation borrowed from \cite{RingLHL}. First, for any $A \in R_{q, n}^{k\times m}$ where $k, m \in \N$:
\begin{align}
    \Lambda^\bot_q(A) \coloneqq \{y\in R^m\vert A\cdot y = 0\mod q\}.
\end{align}
Then, the coefficient embedding $\tau$ takes a polynomial $p\coloneqq \sum_{i\in[n]} p_i X^{i-1}\in R_{q, n}$ and returns a column vector $(p_1, p_2, ..., p_n)^T$. In addition:
\begin{align}
    \Lambda_q(A) \coloneqq \{y\in R^k: \exists s\in R^m \ni A\cdot s=y\mod q\}
\end{align}
Finally, \cite{RingLHL} demonstrates that for $q, n$ such that $R_{q, n}$ is a ring of power-of-$2$ cyclotomic polynomials,
\begin{align}\label{eq:taurot}
    \tau(\Lambda_q^\bot(A))^* &= \Lambda_q^\bot(\mathsf{Rot}(A))^* \\
    &= \frac{1}{q}\Lambda_q(\mathsf{Rot}(A)^T) \\
    &= \frac{1}{q}\{t \in \Z^{kn} : \exists s\in R_{q, n}^{k} \ni t = \mathsf{Rot}(A)^T\cdot \tau(s) \mod q\}.
    \label{eq:existrot}
\end{align}

\begin{lemma}[Properties of $\Rot$]\label{lem:rot-properties}
$\forall a, b \in R_{q, n}$, the following hold:
\begin{align}
    \Rot(a)\tau(b) &= \tau(a b) \\
    \Rot(a)\Rot(b) &= \Rot(a b) \\
    \Rot(a) + \Rot(b) &= \Rot(a + b)
\end{align}
In addition, if there exists $a^{-1} \in R_{q, n}$ such that $a a^{-1} = a^{-1} a = 1$, then:
\begin{align}
    \Rot(a)^{-1} = \Rot(a^{-1})
\end{align}
\end{lemma}
\begin{proof}
    First, let $a, b \in R_{q, n}$ where $a = \sum_{i=0}^{n-1}a_i X^i$ and $b = \sum_{i=0}^{n-1}b_i X^i$. Then, by definition $\Rot(a)\tau(b) = \sum_{j=0}^{n-1}b_j \tau(a X^j).$ Next, using that $\tau$ is additive and respects multiplication by elements of $\Z_q$ we find:
    \begin{align}
        \sum_{j=0}^{n-1}b_j \tau(a X^j) = \tau(a \sum_{j=0}^{n-1}b_jX^j).
    \end{align}
    We may then conclude $\Rot(a)\tau(b) = \tau(a b)$. Next, it immediately follows that:
    \begin{align}
        \Rot(a b) &= [\tau(a b)| ... | \tau(X^{n-1} a b)] = \Rot(a)[\tau(b)| ... | \tau(X^{n-1} b)] = \Rot(a)\Rot(b)
    \end{align}
    To prove additivity, we consider $\Rot(a+b) = [\tau(a+b) | ... | \tau(X^{n-1}(a+b))]$, which by the additivity of $\tau$ is equivalent to:
    \begin{align}
        \Rot(a+b) &= [\tau(a)+\tau(b) | ... | \tau(X^{n-1}a)+\tau(X^{n-1}b)] = \Rot(a) + \Rot(b)
    \end{align}
    Finally, let $a^{-1}$ be the multiplicative inverse of $a$. Then:
    \begin{align}
        \Rot(a)\Rot(a^{-1}) = \Rot(a a^{-1}) = \Rot(1) = I
    \end{align}
\end{proof}
We will now seek to prove the following corollary:
\begin{lemma}\label{lem:keyGenSmooth}
For any $h_B, B\leftarrow \KeyGen$, we find $\eta_\delta(\tau(\Lambda_q^\bot([1|h_B]))) \le \gs{B}\sqrt{\ln(2n(1+1/\delta))/\pi}.$
\end{lemma}
\begin{proof}
First, we consider the smoothing parameter of the lattice $\tau(\Lambda_q^\bot([1|h_B])) = \Lambda_q^\bot(\Rot([1|h_B]^T))$. By definition, our basis matrix is then:
\begin{align}
    \begin{pmatrix}
        \Rot(1) &\Rot(h_B) \\
        0 &\Rot(q)
    \end{pmatrix} = \begin{pmatrix}
        I &\Rot(h_B) \\
        0 &qI
    \end{pmatrix}
\end{align}
Note that from Lemma \ref{lem:rot-properties}, we may conclude:
\begin{align}
    qI = \Rot(fG - gF) = \Rot(f)\Rot(G) - \Rot(g)\Rot(F)
\end{align}
and
\begin{align}
    \Rot(h) = \Rot(g f^{-1}) = \Rot(g)\Rot(f)^{-1}
\end{align}
Using these identities, we can rewrite the basis matrix as follows:
\begin{align}
    \begin{pmatrix}
        I &\Rot(h_B) \\ 0 &qI
    \end{pmatrix} &= \begin{pmatrix}
        \Rot(f) &0 \\ 0 &\Rot(f)^{-1}
    \end{pmatrix}^{-1}\begin{pmatrix}
        \Rot(f) &\Rot(f)\Rot(h_B) \\
        0 &q\Rot(f)^{-1}
    \end{pmatrix}\\
    &= \begin{pmatrix}
        \Rot(f) &0 \\ 0 &\Rot(f)^{-1}
    \end{pmatrix}^{-1}\begin{pmatrix}
        \Rot(f) &\Rot(g) \\
        0 &\Rot(G)-\Rot(g)\Rot(F)\Rot(f)^{-1}
    \end{pmatrix}\\
    &=\begin{pmatrix}
        \Rot(f) &0 \\ 0 &\Rot(f)^{-1}
    \end{pmatrix}^{-1}\begin{pmatrix}
        I &0 \\ -\Rot(F)\Rot(f)^{-1} &I
    \end{pmatrix}\begin{pmatrix}
        \Rot(f) &\Rot(g) \\ \Rot(F) &\Rot(G)
    \end{pmatrix}.
\end{align}
The final matrix on the RHS is a basis for the lattice provided that the preceding transformation is unimodular. That is, we require both of the following matrices to have integer entries and determinant $\pm 1$. By the construction of the $\NTRU$ public key $h_B = gf^{-1}\mod q$, we guarantee that $f$ is invertible and immediately see that both of the following matrices:
\begin{align}
    \begin{pmatrix}
        \Rot(f)^{-1} &0 \\ 0 &\Rot(f)
    \end{pmatrix}, \begin{pmatrix}
        I &0 \\ -\Rot(F)\Rot(f)^{-1} &I
    \end{pmatrix},
\end{align}
have integer elements. Then, being diagonal/lower-triangular, respectively, we see by inspection that the determinant is $1$ in both cases. Therefore, we conclude that the lattice $\tau(\Lambda_q^\bot([1|h_B]))$ admits a basis of the form:
\begin{align}
    \Rot(B) = \begin{pmatrix}
        \Rot(f) &\Rot(g) \\ \Rot(F) &\Rot(G)
    \end{pmatrix}
\end{align}

Now, utilizing Lemma 3.1 from Gentry, Peikert, and Vaikuntanathan, we recognize:
\begin{align}
\eta_\delta(\tau(\Lambda_q^\bot([1|h_B]))) \le \gs{\Rot(B)}\sqrt{\ln(2n(1+1/\delta))/\pi}
\end{align}
for any basis $B$ of the full-rank lattice $\tau(\Lambda_q^\bot([1|h_B]))$ {\cite[Lemma 3.1]{GPV}}.

Now, we claim that $\gs{\Rot(B)} \le \gs{B}$. Let $Q, R$ be an orthogonalization of $B$ such that $B = Q R$ for $Q$ an orthogonal matrix and $R$ an upper triangular matrix. Then, from Definition \ref{def:gsNorm} and the definition of $\Rot(\cdot)$:
\begin{align}
    \gs{B} &= \mathsf{MaxCol}(Q) \\ 
    &= \mathsf{MaxCol}(\Rot(Q)) \\
    &\ge \gs{\Rot(Q)}
    \end{align}
Then, by our Lemma \ref{lem:rot-properties}:
\begin{align}
    \gs{B} &\ge \gs{\Rot(BR^{-1})} \\
    &\ge\gs{\Rot(B)\Rot(R)^{-1}}, \\
    &\ge \gs{\Rot(B)},
\end{align}
with the last inequality clear from our definition of Gram-Schmidt Norm (Def'n \ref{def:gsNorm}). From this claim, it follows that:
\begin{align}   
    \eta_\delta(\tau(\Lambda_q^\bot([1|h_B]))) \le \gs{B}\sqrt{\ln(2n(1+1/\delta))/\pi}.
\end{align}
\qed\vspace{\baselineskip}\end{proof}

This lets us proceed to the next step of proving our main theorem.

\begin{proposition}
    Let $\MRAlg^{H, B_1, B_2}$ be a $\Phi_{H'[\Theta]}$-adversary algorithm with black-box access to $B_1, B_2$. Then, for $\Pr[\Gmeasrep^\MRAlg]$ as defined in Equation \ref{eq:Gmeasrep} and $\Pr[\Gtrapd^\TDAlg]$ in Equation \ref{eq:Gtrapd}:
    \begin{align}
        \abs{\Pr[\Gtrapd^\TDAlg] - \Pr[\Gmeasrep^\MRAlg]} \le  4Q\delta
    \end{align} 
    for some $\delta \in (0, 1/2)$ such that $\sigma \ge 1.1658\sqrt{q}\sqrt{\ln(2n(1+1/\delta))/\pi}$.
    \end{proposition}
\begin{proof}
This distinguishing task is at least as hard as distinguishing between $H$ and $H'_{F}$.
\begin{align}
\begin{split}
    \kappa &\coloneqq \abs{\Pr[\Gtrapd^\TDAlg] - \Pr[\Gmeasrep^\MRAlg]} \\
    &\le \abs{\Pr_{h_B\leftarrow \KeyGen}[b=1\mid b\leftarrow \mathcal{Z}^{H'_{F}, B_1, B_2}(h_B)] - \Pr_{h_B\leftarrow \KeyGen}[b=1\mid b\leftarrow \mathcal{Z}^{H, B_1, B_2}(h_B)]}
\end{split}
\end{align}
for $\mathcal{Z}$ a $2Q$-quantum query distinguishing algorithm with black-box access to $B_1, B_2$. We recognize that this access to $B_1, B_2$ may provide some knowledge of the secret key $B$. Therefore, as an upper-bound, we consider the case where $\mathcal{Z}$ has complete access of $B$ for this distinguishing task:
\begin{align}
\begin{split}
    \kappa &\le \abs{\Pr_{h_B\leftarrow \KeyGen}[b=1\mid b\leftarrow \mathcal{Z}^{H'_{F}}(h_B, B)] - \Pr_{h_B\leftarrow \KeyGen}[b=1\mid b\leftarrow \mathcal{Z}^{H}(h_B, B)]}
\end{split}
\end{align}

Next, we concern ourselves with the following quantity for $\mathcal{Z}^\mathcal{O}$ when allowed exactly $j$ queries to the oracle $\mathcal{O}$:
\begin{align}
    \Adv^{\mathsf{Dist}}_j(\mathcal{Z}) \coloneqq \abs{\Pr_{h_B\sample R_{q, n}}[b=1\mid b\leftarrow \mathcal{Z}^{H'_{F}}(h_B, B)] - \Pr_{h_B\sample R_{q, n}}[b=1\mid b\leftarrow \mathcal{Z}^{H}(h_B, B)]}
\end{align}
Now, we make a hybrid argument acting one additional query from $H$ (resp. $H'_F$) for each round. By the triangle inequality, it follows that:
\begin{align}
    \Adv^{\mathsf{Dist}}_{2Q}(\mathcal{Z}) \le 2Q\Adv^{\mathsf{Dist}}_{1}(\mathcal{Z})
\end{align}
With this established, we now need only consider the following for $\ket{\psi} \coloneqq \sum_{r\in \{0, 1\}^k}p_r\ket{r}\ket{0}$ an arbitrary pure state:
\begin{align}
\begin{split}
    \mathrm{Adv}^{\mathrm{Dist}}_1(\mathcal{Z}) &= \abs{\Pr[b=1\vert b\leftarrow \mathcal{Z}(h_B, B, \ket{\psi}, H'\ket{\psi})]-\Pr[b=1\vert b\leftarrow \mathcal{Z}(h_B, B, \ket{\psi}, H\ket{\psi})]} \\
    &\le \mathsf{TD}\left(H'\ket{\psi}, H\ket{\psi}\right)
\end{split}
\end{align}
where $\mathsf{TD}$ denotes the trace distance. Note that for a pure state (or any commuting density matrix mixed states), trace distance is equivalent to the total variational distance (aka statistical distance) of the corresponding probability distributions. Thus, we now apply Lemma \ref{lem:latticeSmooth} and Lemma \ref{lem:keyGenSmooth} to find:
\begin{align}
    \mathsf{TD}\left(H'\ket{\psi}, H\ket{\psi}\right) \le 2\delta
\end{align}
for $\sigma \ge \gs{B}\sqrt{\ln(2n(1+1/\delta))}$. Recall $\gs{B} \le 1.1658\sqrt{q}$ for our analysis. Then, the proposition follows under the application of Bernoulli's inequality.
\qed\vspace{\baselineskip}\end{proof}

Our newest algorithm, $\TDAlg^{H'_{F}, B_1, B_2}$, can be written as acting on some initial state $\phi_0$ with reprogramming at the $i$th query as follows:
\begin{align}
    \TDAlg^{H'_{F}[s^*], B_1, B_2}\ket{\phi_0} = R_{2Q} O^{\ket{H'_{F}[s^*]}}R_{2Q-1}...R_{i+2} O^{\ket{H'_{F}[s^*]}}R_{i+1} O^{s^*}XR_{i}O^{\ket{H'_{F}}}R_{i-1} ...R_1 O^{\ket{H}}\ket{\phi_0}
\end{align}
for $s^* \coloneqq (s_1^*, s_2^*)$ with $s_1^*, s_2^*$ selected according to Gaussians over $R_{q, n}$ with standard deviation $\sigma$. Here, $O^{s^*}$, the query made immediately after a complete measurement of the query register returns some $x_0 \in X$, has a $0.5$ chance of responding with $H'_{F}(x_0)$ and a $0.5$ chance of responding with $H'_{F}[s^*](x_0)$, the latter of which is - by construction - equal to $[1|h_B]\cdot s^* \mod q$.

As before, the above - while less than elegant - can be understood straightforwardly as a quantum algorithm outputting a solution utilizing a single classical query (at our reprogrammed location) to an oracle from the family $\mathsf{F}_{h_B}$ which returns $[1|h_B]\cdot s^* \mod q$ in response to the query $x_0$. All other operations are completed with no knowledge of this 'true' oracle and can be expressed more succinctly - though less informatively - as:
\begin{align}
    \TDAlg^{H'_{F}[s^*], B_1, B_2}\ket{\phi_0} = U_2 O^{s^*}XU_1\ket{\phi_0}.
\end{align}
for some unitaries $U_1, U_2$. Now, in order to avoid our notation becoming any more cumbersome from here, we'll let:
\begin{align}
    H^*(r) &\coloneqq H'_{F}[s^*](r), \\
    F'(r) &\coloneqq \begin{cases}
         s^* &r=x_0 \\
         F(r) &r\ne x_0
    \end{cases}
\end{align}
for all $r \in \{0, 1\}^k$.

Now, we replace access to $B_1, B_2$ with access to black-boxes $B_1', B_2'$. This is our first step towards replacing $B_1, B_2$ with operations that do not require access to the secret key $B$. Let $B_1'$ simply select $r$ uniformly at random from $\{0, 1\}^k$; this is the same as $B_1$. More importantly, we define $B_2'$ as:
\begin{enumerate}
    \item Given $r \in \{0, 1\}^k$
    \item $s_2 \coloneqq \bot$
    \item While $s_2 = \bot$:
    \begin{enumerate}
        \item $(s_1, s_2) \leftarrow D_{\Lambda', \sigma, 0}(u)$ where $\Lambda'$ is the shifted lattice $\Lambda(B)_{(c, 0)}$, $u$ is a random seed, and $D_{\Lambda, \sigma, \mu}(u)$ denotes a deterministic construction for $D_{\Lambda, \sigma, \mu}$ and any fixed $u$
        \item If $\vert (s_1, s_2) \vert^2 > \beta^2$, $s_2 \coloneqq \bot$.
    \end{enumerate}
    \item Return $(r, s_2)$
\end{enumerate}
Accordingly, this produces our next experiment.

\paragraph{Game $\GRepOne$} This is the same as $\Gtrapd$ except we replace $B_1, B_2$ with the previously defined $B_1', B_2'$:
\begin{align}
    \GRepOne \coloneqq \Pr_{pk, sk \leftarrow \KeyGen}[\Verify(pk, m, \sigma), \ m \notin \textup{SignQ} \mid (m, \sigma)\leftarrow\ROAlg^{H^*}].\label{eq:GRepOne}
\end{align}

\begin{lemma}
    Let $\TDAlg$ (resp. $\ROAlg$) be $\Phi_{H^*}$-adversary algorithms with up to $Q$ black box accesses to $B_1, B_2$ (resp. $B_1', B_2'$). Then, for $\Pr[\GRepOne^\ROAlg]$ as defined in Equation \ref{eq:GRepOne} and $\Pr[\Gtrapd^\TDAlg]$ from Equation \ref{eq:Gtrapd}:
    \begin{align}
        \abs{\Pr[\GRepOne^\ROAlg] - \Pr[\Gtrapd^\TDAlg]} \le Q2^{-\lambda},
    \end{align}
    provided that for $\lambda \in \N, \zeta \coloneqq 2^{-\lambda}/4n$, $\sigma \ge \eta_\zeta(\Z)1.1658\sqrt{q}$, see Eq'n \ref{eq:ks-sd}.
\end{lemma}

\begin{proof}
    This is indistinguishable provided that the output of $\Klein$ is indistinguishable from $s_1, s_2$ selected according to $D_{\Lambda', \sigma, 0}$. As otherwise, $B_2$ and $B_2'$ are step-for-step identical. Then, by Equation \ref{eq:ks-sd} and Bernoulli's inequality to account for up to $Q$ successive queries to $\Sign_H$, the lemma immediately follows. Recall $\gs{B} \le 1.1658\sqrt{q}$.
\qed\vspace{\baselineskip}\end{proof}

Next, we again replace $B_2'$ with a new black-box $B_2''$ now modified as to have no access to the secret key $B$. We define $B_2''$ as:
\begin{enumerate}
    \item Given $r \in \{0, 1\}^k$
    \item $(s_1, s_2) \coloneqq F'(r)$
    \item If $\vert (s_1, s_2) \vert^2 > \beta^2$, return $\bot$
    \item Otherwise, return $(r, s_2)$
\end{enumerate}
Similarly, we define $B_1''$ as selecting $r\sample \{r\in \{0, 1\}^k|( s_1, s_2) \coloneqq F'(r), \abs{s_1, s_2}^2\le\beta^2\}$. This lets us define the following experiment:

\paragraph{Game $\GRepTwo$} This is the same as $\GRepOne$ except we replace $B_1'$ (resp. $B_2'$) with the previously defined $B_1''$ (resp. $B_2''$):
\begin{align}
    \GRepTwo \coloneqq \Pr_{pk, sk \leftarrow \KeyGen}[\Verify(pk, m, \sigma), \ m \notin \textup{SignQ} \mid (m, \sigma)\leftarrow\RTAlg^{H^*, B_1'', B_2''}(pk)].\label{eq:GRepTwo}
\end{align}

\begin{lemma}
    Let $\ROAlg$ (resp. $\RTAlg$) be $\Phi_{H^*}$-adversary algorithms with black-box access to $B_1', B_2'$ (resp. $B_1'', B_2''$). Then, for $\Pr[\GRepTwo^\RTAlg]$ as defined in Equation \ref{eq:GRepTwo} and $\Pr[\GRepOne^\ROAlg]$ from Equation \ref{eq:GRepOne}:
    \begin{align}
        \abs{\Pr[\GRepTwo^\RTAlg] - \Pr[\GRepOne^\ROAlg]} = 0 
    \end{align}
\end{lemma}
\begin{proof}
This is indistinguishable provided that $B_2''$ does not output $\bot$, as by construction we know the outputs of $F'$ are indistinguishable from the results selected according to $D_{\Lambda', \sigma, 0}$. Then, by design of $B_1''$, we guarantee that $B_2''$ will never output $\bot$ and the result follows
\qed\vspace{\baselineskip}\end{proof}

At this point $B_1''$ and $B_2''$ can both be completed without any access to the secret key $B$. Thus, $\GRepTwo$ is immediately equivalent to breaking the $\UFNMA$ security of $\FALCON^{(+)}$ from a single reprogrammed query, knowledge of the public key, and no further resources. As a result, the challenge of passing $\GRepTwo$ with an algorithm $\ISAlg$ can be captured by the following:
\begin{align}
    \Pr[\GRepTwo'^\ISAlg] \coloneqq \Pr_{h\leftarrow \mathsf{KeyGen}}\Bigl[[1|h]\cdot y = t \mod q \wedge \left|  y \right|^2\le \beta^2\mid t \leftarrow R_{q, n}, \ y \leftarrow \ISAlg (h, t) \Bigr]
\end{align}
We use the $\GRepTwo'$ notation to formalize the knowledge that given any $\GRepTwo$-adversary $\RTAlg$, we may construct an algorithm $\ISAlg$ that implements the interaction of $\RTAlg$ using $B_1''$ and $B_2''$ implicitly.

\paragraph{Game $\GRISIS$:}\label{sec:g5} The Hermite-form of the lattice problem $\RISIS$. 
\begin{align}
\begin{split}
   \Pr[\GRISIS^{\ISAlg}] &\coloneqq \Adv^{\RISIS}_{1, \beta}(\ISAlg), \\
   &= \Pr_{h\leftarrow R_{q, n}}\Bigl[[1|h]\cdot y = t \mod q \wedge \abs{y}^2\le \beta^2\mid t\sample R_{q, n}, \ y \leftarrow\ISAlg(h, t) \Bigr].\label{eq:GRISIS}
\end{split}
\end{align}

Finally, we may recognize that our algorithm breaking the $\UFNMA$ security of $\FALCON^{(+)}$ at an adaptively reprogrammed oracle location is at least as hard as solving the average-case Hermite-form $\RISIS$ problem. 
\begin{lemma}
    Let $\RTAlg$ be an algorithm with running time $T$ and $\calN$ an algorithm with running time no more than $T$. Then,
    \begin{align}
        \Pr[\GRISIS^{\ISAlg}] + \Adv^{\NTRU}_{\sigma_{f, g}}(\calN) &\geq p_{GS}\cdot \Pr[\GRepTwo'^\ISAlg]
    \end{align}
\end{lemma}
\begin{proof}
    Let $\mathbb{D} \coloneqq \{h\coloneqq gf^{-1}\mod q: f, g \leftarrow D_{R_{q, n}^\times, \sigma_{f, g}, 0} \times D_{R_{q, n}, \sigma_{f, g}, 0}\}$ be a distribution over $R_{q, n}$. Now, observe by construction that:
    \begin{align}\label{eq:claim1}
        \Pr_{h\leftarrow \mathbb{D}}\Bigl[[1|h]\cdot y = t \mod q \wedge \abs{y}^2\le \beta^2\mid t\sample R_{q, n}, \ y \leftarrow\ISAlg(h, t) \Bigr] \ge p_{GS}\Pr[\GRepTwo'^\ISAlg].
    \end{align}
    That is, the probability that $\ISAlg$ successfully computes a valid $y$ under the NTRU public key distribution is at least $p_{GS}$ times the probability that $\ISAlg$ successfully computes a valid $y$ when the Gram–Schmidt bound is additionally imposed.
    
    Next, we recognize that distinguishing between $h_B$ chosen according to $\mathbb{D}$ and $h_B$ chosen uniformly at random from over $R_{q, n}$ within our parameter regime is equivalent to the $\NTRU$ learning problem as defined in Eq'n \ref{eq:ntru}. To incorporate this, we again resolve the identity:
    \begin{align}
    \begin{split}
        \Pr[\GRISIS^\ISAlg]&= \Pr_{h\leftarrow R_{q, n}}\Bigl[[1|h]\cdot y = t \mod q \wedge \abs{y}^2\le \beta^2\mid t\sample R_{q, n}, \ y \leftarrow\ISAlg(h, t) \Bigr]\\
        &= \Pr_{h\leftarrow R_{q, n}}\Bigl[[1|h]\cdot y = t \mod q \wedge \abs{y}^2\le \beta^2\mid t\sample R_{q, n}, \ y \leftarrow\ISAlg(h, t) \Bigr] \\ &\qquad\qquad- \Pr_{h\leftarrow \mathbb{D}}\Bigl[[1|h]\cdot y = t \mod q \wedge \abs{y}^2\le \beta^2\mid t\sample R_{q, n}, \ y \leftarrow\ISAlg(h, t) \Bigr] \\&\qquad\qquad+ \Pr_{h\leftarrow \mathbb{D}}\Bigl[[1|h]\cdot y = t \mod q \wedge \abs{y}^2\le \beta^2\mid t\sample R_{q, n}, \ y \leftarrow\ISAlg(h, t) \Bigr] \\
        &\ge \Pr_{h\leftarrow \mathbb{D}}\Bigl[[1|h]\cdot y = t \mod q \wedge \abs{y}^2\le \beta^2\mid t\sample R_{q, n}, \ y \leftarrow\ISAlg(h, t) \Bigr] - \Adv^\NTRU_{\sigma_{f, g}}(\calN)
    \end{split}
    \end{align}
    Here, we construct an algorithm $\calN$ that, given a challenge public key $h$, samples $t\sample R_{q, n}$ and runs $\ISAlg(h,t)$ to obtain $y$. It outputs $1$ if and only if $[1|h]\cdot y=t\mod q$ and $\abs^2\leq\beta^2$. This results in a running time no larger than $O(T)$ for $T$ the running time of $\ISAlg$. Combining this with Eq'n \ref{eq:claim1} we find:
    \begin{align}
        \Pr[\GRISIS^{\ISAlg}] + \Adv^{\NTRU}_{\sigma_{f, g}}(\calN) &\geq p_{GS}\cdot \Pr[\GRepTwo'^\ISAlg]
    \end{align}
    as previously claimed.
\qed\vspace{\baselineskip}\end{proof}

The proof for the theorem now follows directly by combining the lemmas proven above. Particularly, we may write for $\SRAlg$ any polynomial quantum query algorithm and $\calN, \ISAlg$ any algorithms with access to $\SRAlg$:
\begin{align}
\begin{split}
    \Adv&_{\FALCON^+}^{Q-\UFCMA}(\UFAlg) = \Pr[\Gufcma^\UFAlg] \\
     &\le  \Pr[\Gsignrep^\SRAlg] + \frac{Q}{2}\frac{\chi + 2^{-\lambda}}{1-\chi} \\
     &\le  (4Q+1)^2\left(\Pr[\Gmeasrep^\MRAlg]\right) + \frac{Q}{2}\frac{\chi + 2^{-\lambda}}{1-\chi} \\
     &\le (4Q+1)^2\left(\Pr[\Gtrapd^\TDAlg] + 4Q\delta\right) + \frac{Q}{2}\frac{\chi + 2^{-\lambda}}{1-\chi} \\
     &\le (4Q+1)^2\left(\Pr[\GRepOne^\ROAlg] + 4Q\delta + Q2^{-\lambda}\right) + \frac{Q}{2}\frac{\chi + 2^{-\lambda}}{1-\chi} \\
     &\le (4Q+1)^2\left(\Pr[\GRepTwo^\RTAlg] + 4Q\delta + Q2^{-\lambda}\right) + \frac{Q}{2}\frac{\chi + 2^{-\lambda}}{1-\chi} \\
     &\le (4Q+1)^2p_{GS}^{-1}\left(\Adv_{\sigma_{f, g}}^{\NTRU}(\calN) + \Adv_{1, \beta}^{\RISIS}(\ISAlg)\right) + \frac{Q}{2}\frac{\chi + 2^{-\lambda}}{1-\chi} + (4Q+1)^2\left(4Q\delta + Q2^{-\lambda}\right)
\end{split}
\end{align}
Our theorem rigorously proves that breaking the $\UFCMA$ security of $\FALCON^{(+)}$ requires breaking either the $\NTRU$ Learning or the Hermite-form $\RISIS$ problem. In addition, we will now show that performing key recovery on $\FALCON^{(+)}$ and the difference between breaking the $\sUFCMA$ and $\UFCMA$ securities of $\FALCON^{(+)}$ are both no easier than breaking the $\UFCMA$ security directly. As a result, our final concrete analysis will focus on the $\UFCMA$ security. This marks the end of our formal results.

\subsection{Other security results}
\subsubsection{Key Recovery}
To convince ourselves that proving $\UFCMA$ security is sufficient for $\FALCON^{(+)}$ we consider what would be needed for an adversary to perform key recovery against $\FALCON^{(+)}$.

An adversary completes a key recovery attack on $\FALCON^{(+)}$ if given some $h_B$, $\calA$ can solve for $f, g$ such that $h_B = gf^{-1}\mod q$. This is by definition equivalent to the $\NTRU$ learning problem such that for any efficient quantum algorithm $\calA$ there exists an efficient $\calB$ such that:
\begin{align}
    \Adv^{\mathsf{Key-Recovery}}_{\FALCON^{(+)}}(\calA) \le \Adv^{\NTRU}_{\sigma_{f, g}}(\calB).
\end{align}
Then, as ensuring $\UFCMA$ security for $\FALCON^{(+)}$ -as discussed in Section \ref{sec:Games}- requires that the quantity $(4Q+1)^2\Adv^{\NTRU}_{\sigma_{f, g}}(\calB)$ be upper-bounded by $2^{-b}$ for the desired bit security $b$, it follows that:
\begin{align}
    \Adv^{\mathsf{Key-Recovery}}_{\FALCON^{(+)}}(\calA) \le \Adv^{\NTRU}_{\sigma_{f, g}}(\calB) \le (4Q+1)^2\Adv^{\NTRU}_{\sigma_{f, g}}(\calB) \le 2^{-b}.
\end{align}
Thus, for any nonzero $Q$, $\UFCMA$ security implies security against any efficient key recovery attacks. Accordingly, our concrete analysis will focus on $\UFCMA$ attacks instead of key recovery attacks because for any $b$ bit security that we guarantee for $\UFCMA$, it follows that $\FALCON^{(+)}$'s bit security against key recovery attacks must be even greater than $b$.

\subsubsection{Strong Unforgeability}
For completeness, we may also consider the difference between strong unforgeability and unforgeability against chosen message attacks on $\FALCON^+$:
\begin{align}
    \zeta &\coloneqq \abs{\Adv^{\sUFCMA}_{\FALCON^+}(\calA) - \Adv^{\UFCMA}_{\FALCON^+}(\calB)}, \\
\begin{split}
    &= \abs{\Pr_{pk, sk\leftarrow\KeyGen}[\Verify_{pk}(m, \sigma) \wedge \ (m, \sigma) \notin \textup{SignQR} \mid (m, \sigma)\leftarrow\calA^{\ket{H}, \Sign_H(sk, \cdot)}(pk)] \\
    &\qquad\qquad- \Pr_{pk, sk\leftarrow\KeyGen}[\Verify_{pk}(m, \sigma)  \wedge \ m \notin \textup{SignQ} \mid (m, \sigma)\leftarrow\calA^{\ket{H}, \Sign_H(sk, \cdot)}(pk)]}.
\end{split}
\end{align}
where $\textup{SignQ}$ includes all messages $m$ sent to $\Sign_H$ while $\textup{SignQR}$ includes all message and signature pairs $(m, \sigma)$ produced by $\Sign_H$ as described in Definitions \ref{def:sUF} and \ref{def:UF}. Accordingly, we may write:
\begin{align}
\begin{split}
    \Adv^{\sUFCMA}_{\FALCON^{(+)}}(\calA) &= \Pr_{\substack{pk, sk\leftarrow\KeyGen\\ (m, \sigma)\leftarrow\calA^{\ket{H}, \Sign_H(sk, \cdot)}(pk)}}[\Verify_{pk}(m, \sigma) \wedge \ m \notin \textup{SignR} \wedge (m, \sigma) \notin \textup{SignQR}] \\
    &\qquad\qquad+ \Pr_{\substack{pk, sk\leftarrow\KeyGen\\ (m, \sigma)\leftarrow\calA^{\ket{H}, \Sign_H(sk, \cdot)}(pk)}}[\Verify_{pk}(m, \sigma) \wedge \ m \in \textup{SignR} \wedge (m, \sigma) \notin \textup{SignQR}],
\end{split} \\
\begin{split}
    &= \Adv^{\UFCMA}_{\FALCON^{(+)}}(\calA) \\
    &\qquad\qquad+ \Pr_{\substack{pk, sk\leftarrow\KeyGen\\ (m, \sigma)\leftarrow\calA^{\ket{H}, \Sign_H(sk, \cdot)}(pk)}}[\Verify_{pk}(m, \sigma) \wedge \ m \in \textup{SignR} \wedge (m, \sigma) \notin \textup{SignQR}].
\end{split}
\end{align}
Now, we may assume that no message or public key are easier to forge signatures for than others. Then we may write:
\begin{align}
    \zeta &\le 1 - (1-\Pr_{pk, m}[\Verify_{pk}(m, \sigma') \wedge \sigma \ne \sigma'\mid \sigma \coloneqq \Sign_{H, pk}(m),\sigma'\leftarrow\calA^{\ket{H}}(m, \sigma)])^Q.
\end{align}
Incorporating the behavior of $\Verify$ and recalling that $H$ is a random oracle, we find:
\begin{align}
    \zeta &\le 1 - (1-\Pr_{\substack{pk \leftarrow \KeyGen, \\(r, s_2) \coloneqq \Sign_H}}[\abs{(s_2 - s_2')h \mod q}^2 \le \beta^2\wedge s_2 \ne s_2'\mid s_2'\leftarrow\calA^{\ket{H}}(s_2)])^Q.
\end{align}
We recognize the probability above as $\Adv^{\RISIS}_{1, \beta}$. Thus, for any quantum algorithm $\calA$ we may write: 
\begin{align}
    \zeta \le 1-(1-\Adv^{\RISIS}_{1, \beta}(\calA))^Q.
\end{align}
Then, applying Bernoulli's inequality, we may write:
\begin{align}
    \zeta \le Q\Adv^{\RISIS}_{1, \beta}(\calA)
\end{align}
Finally, as ensuring $\UFCMA$ security for $\FALCON^{(+)}$ -as discussed in Section \ref{sec:Games}- requires that the quantity $(4Q+1)^2\Adv^{\RISIS}_{1, \beta}(\calA)$ be upper-bounded by $2^{-b}$ for the desired bit security $b$, it follows that:
\begin{align}
\begin{split}
    \Adv^{\sUFCMA}_{\FALCON^{(+)}}(\calA) &\le \Adv^{\UFCMA}_{\FALCON^+}(\calB) + Q\Adv^{\RISIS}_{1, \beta}(\calC) \\
    &\le \Adv^{\UFCMA}_{\FALCON^{(+)}}(\calB) + (4Q+1)^2\Adv^{\RISIS}_{1, \beta}(\calC) \\
    &\le \Adv^{\UFCMA}_{\FALCON^{(+)}}(\calB) + 2^{-b}
\end{split}
\end{align}
for any quantum algorithms $\calA, \calB, \calC$ where our parameters guarantee $\UFCMA$ security. Thus, $b$ bit $\UFCMA$ security implies $b$ bit $\sUFCMA$ security.

\subsection{Resulting concrete analysis}\label{sec:concrete}
Now we seek to quantify the concrete security of $\FALCON^{(+)}$ under the security reduction proven above. These results will rely on some heuristics throughout, but are as rigorous as current techniques allow. First, we quantify the hardness of $\RISIS$ against quantum algorithms.

\subsubsection{Hardness of $\RISIS$}\label{sec:risis}
While $\RISIS$ is no harder than $\ISIS$ by definition, there exists no known method for taking advantage of the ring structure. Therefore, a conventional heuristic assumes that the two problems are equally hard. More specifically, we will \textbf{assume}:
\begin{align}
    \Adv^{\RISIS}_{k, m, \beta}(\calA) = \Adv^{\ISIS}_{kn, m n, \beta}(\calA).
\end{align}
for $k, m, \beta \in \N$.

Next, we can show that $\ISIS$ and $\SIS$ are similarly hard.
\begin{lemma}\label{lem:sis-isis}
Let $\calB$ be a quantum algorithm and $k, m, \beta \in \N$. Then:
\begin{align}
     \Adv^{\SIS}_{k, m+1, \beta}(\calB) \ge \Adv^{\ISIS}_{k, m, \beta}(\calB) \ge \frac{1}{2\beta(m+1)}\Adv^{\SIS}_{k, m+1, \beta}(\calB).
\end{align}
\end{lemma}
\begin{proof}
For the first inequality, let $\calA$ be an algorithm that attempts to solve the $\ISIS$ problem such that for $k, m, \beta \in \N$, $\calA$ has success probability:
\begin{align}
    \sigma_{\calA} \coloneqq \Pr[A\cdot x = t \mod q \wedge \abs{x} \le \beta|A \sample \Z_q^{k\times m}, t \sample \Z_q^k, x \leftarrow \calA_{q, k, m, \beta}(A, t)].
\end{align}
Then we may define the following algorithm $\calB_{k, m, \beta}(A)$ that solves the $\SIS$ problem for any given operator $A \sample \Z_q^{k\times m}$:
\begin{enumerate}
    \item For any $j \in [m]$, parse $A = [A_1 | A_2]$ as $A_1 \in \Z_q^{k\times j}, A_2 \in \Z_q^{k\times (m-j)}$
    \item Select $e_1 \sample \{v\in \Z_q^j: \abs{v}\le\beta\}$
    \item $y \coloneqq -A_1 \cdot e_1 \mod q$
    \item $e_2 \leftarrow \calA(A_2, y)$
    \item Return $(e_1, e_2)$
\end{enumerate}
$\calB$ is successful provided that $\calA$ is successful when called in Step 4. For $A$ distributed uniformly over $\Z_q^{k\times m}$ and nonzero $e_1$, it follows that $A_2$ (resp. $A_1 e_1$) is distributed uniformly over $\Z_q^{k\times (m-j)}$ (resp. $\Z_q^{k\times j}$). Therefore, $\calB$ has the same success probability $\sigma_{\calA}$ as $\calA$. We conclude that for any efficient algorithm solving $\ISIS$ with nonzero probability, then there exists an efficient algorithm solving $\SIS$ with the same odds of success.

For the second inequality, given an algorithm $\calB$ that solves the $\SIS$ problem and a given $\ISIS$ operator $A\in \Z_q^{k\times m}$ and target vector $t \in \Z_q^k$, we select a random $j \sample [m]$ and for $A_i$ the $i$th column vector of $A$ we may define:
\begin{align}
    A' \coloneqq [A_1|A_2|...|A_{j-1}|t|A_j|...|A_m].
\end{align}
Then, we apply $\calB$ to this operator $A'$ returning a small vector $x \in \Z_q^{m +1}$. This provides a solution to the $\ISIS$ problem if and only if the $j$th element of $x$ is equal to $-1$ as this element's location corresponds to the target vector's insertion. As a result, a more optimal strategy is to instead select $z \sample [[-\gamma, \gamma]] \setminus \{0\}$, $j \sample [m]$ then define:
\begin{align}
    A'' \coloneqq [A_1|A_2|...|A_{j-1}|\frac{t}{z} \mod q|A_j|...|A_m].
\end{align}
Now when we apply $\calB$ to $A''$ to return a small vector $x'$ we can expect $( x_1, ..., x_{j-1}, x_{j+1}, ..., x_{m})$ to constitute a solution to the given $\ISIS$ instance if and only if $x_j$ is equal to $-z$. We must take into account the probability that $z$ is nonzero as well, so we find:
\begin{align}
    \Pr[x_j = -z] &= \sum_{a \in [[-\gamma, \gamma]]\setminus \{0\}}\Pr[x_j=a]\Pr[z=-a] \\
    &= \frac{1}{2\gamma}\Pr[x_j\ne 0] \ge \frac{1}{2\gamma(m+1)}
\end{align}
Now by definition, any coefficient for a valid solution to the $\SIS$ problem will lie within the some subset such that the full solution lies in $S_\beta \coloneqq \{\abs{v}\le\beta: v\in \Z_q^n\} \subseteq \Z_q^n$. Thus, if we let $\gamma \in [0, q]$ be the largest value that a coefficient may be while still allowing the vector to lie in $S_\beta$ for bound $\beta$ and norm $\abs{\cdot}$, it follows that this approach succeeds with non-negligible probability like $1/2\gamma(m+1)$ times the success probability of $\calB$.
\qed\vspace{\baselineskip}\end{proof}

Finally, following \cite{Lyubashevsky12}, we estimate the security of $\SIS_{k, m,\beta}$, by considering the attack that uses the Block Korkine-Zolotarev ($\BKZ$) algorithm and following the Core-SVP methodology introduced in \cite{AlkimNewHope2016}. By this method, we assume the optimal attack on the $\SIS$ problem is the Block Khorkine-Zolotarev (BKZ) algorithm, which makes calls to an SVP solver for $\mu$-dimension lattices. Then, the method makes a conservative estimate of the run-time of the $\BKZ$ algorithm as equal to the run-time of a single call to the solver. This cost is approximated $2^{0.265\mu}$ as this is the cost of the best known quantum $\SVP$ solver from Laarhoven \cite{Laarhoven16}. To estimate the block-size $\mu$, we define the \textit{root Hermite factor} $\delta\colon \mathbb{N} \to \mathbb{R}$:
\begin{equation}\label{eq:rootHerm}
   \delta(\mu) \coloneqq \Bigl(\frac{(\mu\pi)^{1/\mu} \mu}{2\pi e}\Bigr)^{\frac{1}{2(\mu-1)}}.
\end{equation}
Then, to solve some instance of $\SIS_{k,m,\beta}$, we set the BKZ block-size $\mu$ to the smallest positive integer such that
\begin{equation}
    \cdot  2^{2 \sqrt{k \log(q) \log(\delta(\mu)) }} \leq \beta.
\end{equation}

We may now move on to quantifying the hardness of the $\NTRU$ learning problem against quantum algorithms.

\subsubsection{Hardness of NTRU Learning}
The best known algorithm for distinguishing our public key $h = gf^{-1}\mod q$ from a polynomial $h$ selected uniformly at random from $R_{q, n}$ requires finding unusually small $f, g$ such that:
\begin{align}
    h f - g = 0 \mod q.
\end{align}
Then, solving this problem is equivalent to locating the vector $(f, g)$ in a $2n$-dimensional lattice with determinant $q^n$ \cite{Ducas14}. Experimental results suggest that finding any unusually small vector $v$ in an NTRU lattice $\Lambda$ of dimension $N$ results in a root Hermite factor, $\delta$, such that \cite{Ducas13}:
\begin{align}
    0.4\delta_0^N = \frac{\sqrt{N/(2\pi e)}\det(\Lambda)^{1/N}}{\abs{v}}.
\end{align}
Now applying the $\BKZ$ algorithm to this lattice problem, we estimate the required BKZ block size $\mu$ as the smallest integer for which the root Hermite factor achieved by BKZ is no greater than the target root Hermite factor $\delta_0$.
for $\delta(\mu)$ as defined in Eq'n \ref{eq:rootHerm}.

\subsubsection{Parameter Sets}
Now, we present parameters that allow for $\FALCON^{(+)}$'s security under our reduction. This signature scheme is parameterized by the following five variables:
\begin{align}
    Q, q, n, k, \tau_0.
\end{align}
Here, $Q$ gives the upper-bound on the number of queries to the hash function allowed to an adversarial algorithm. In order to satisfy the bit security requirements for NIST security level $l\in [5]$. Conservatively, we'll upper bound $Q$ by $2^c$, for $c$ as described in the table below:
\begin{table}[ht]
\centering\renewcommand{\arraystretch}{1.5}
    \begin{tabular}{|c|p{1cm}|p{1cm}|p{1cm}|p{1cm}|}
        \hline
        NIST Security Level (SL$\ell$) &  SL1 & SL2 & SL3 & SL4/5\\
        \hline
        Upper bound on $Q$ $(C_\ell)$ &  ${64}$ & ${86}$ & ${96}$ & ${128}$ \\
        \hline
    \end{tabular}
    \caption{Upper bounds on $Q$ for NIST security levels 1 to 5. These numbers are based on \cite[Appendix A]{FIPS204} combined with the best-known quantum query complexity results when block ciphers and hash functions are modeled as random functions.}
    \label{tab:nist_level_query}
\end{table}
For $q$ we require a prime number and $n$ a power of $2$ such that $q \equiv 1\mod 2n$. Finally, $k$ gives the size of the random seed $r \sample \{0, 1\}^k$ which makes up part of every signature. As our analysis is in the QROM, $k$ has no impact on the security results stemming from our analysis, but in practice it should be chosen such that the hash function $H(.)$ is "close enough" to a truly random function. As the original design for $\FALCON^{(+)}$ uses $k=320$, we will do the same \cite{FALCON}.

From \ref{eq:theo} and our analysis in Section \ref{sec:risis}, we conclude that to assure ourselves that $\FALCON^{(+)}$ has some bit security $c$, we require that all of the following quantities are less than $2^{-c}$ for any efficient quantum algorithms $\calN, \ISAlg$:
\begin{itemize}
    \item $(4Q+1)^2p_{GS}^{-1}\left(\Adv_{\sigma_{f, g}}^{\NTRU}(\calN)\right)$
    \item $(4Q+1)^2p_{GS}^{-1}\left(\Adv_{1, \beta}^{\RISIS}(\ISAlg)\right)$ \\
    \item $Q\left(\frac{\chi + 2^{-\lambda}}{2-2\chi}\right)$, $Q(4Q+1)^2(4\delta)$, $Q(4Q+1)^2(2^{-\lambda})$
\end{itemize}
Here, for $\beta, \lambda\in \N$, $\zeta \coloneqq 2^{-\lambda}/4n$, and $\sigma \ge \max\left(\eta_\zeta(\Z), \sqrt{\ln(2n(1+1/\delta))/\pi}\right)1.1658\sqrt{q}$.

For the non-computational term(s), we require:
\begin{align}
    (2^c)(2^{c+2})^2(2^{-\lambda}) \le 2^{-c}&\implies \lambda \ge 4c+4, \\
    (2^c)(2^{c+2})^2(2^2\delta) \le 2^{-c} &\implies \delta \le 2^{-4c-6}.
\end{align}
Within our parameter regime, the contribution of the $2^{-2n}$ term to $\chi$ is negligible. We therefore fix $\epsilon \coloneqq 1/8$, yielding
\begin{align}\label{eq:chi}
\chi = 2^{-\lambda}+\frac{9}{7}2^{-2n}
=2^{-\lambda}+\operatorname{negl}(n).
\end{align}
We can then confirm that this suffices for our security conditions as for any $\chi \in (0, 1/2)$:
\begin{align}
(2^c)\left(\frac{\chi+2^{-\lambda}}{2-2\chi}\right)
\le 2^c\left(\chi+2^{-\lambda}\right).
\end{align}
Thus, it suffices to require $\chi \le 2^{-2c}-2^{-\lambda}$.
Since $\lambda=4c+4$, this is implied by $\chi\le 2^{-2c-1}$, which is satisfied by Eq.~\ref{eq:chi} within our parameter regime.

Finally, we observe that our parameter choices allow us to consolidate the three lower bounds on $\sigma$. If we fix $\lambda\coloneqq4c+4$ and $\delta\coloneqq2^{-4c-6}$, we have
\begin{align}
\ln\left(2n(1+1/\delta)\right)
&\le \ln(8n2^{4c+4}) =\ln(2/\zeta).
\end{align}
Consequently, by the standard lower bound on the smoothing parameter, $\eta_\zeta(\mathbb Z) \ge \sqrt{\ln(2/\zeta)/\pi}$, the second lower bound is implied by the first. Furthermore, since $\epsilon=1/8$ and $\lambda=4c+4$, we have $\epsilon/n\ge\zeta$, and hence, by monotonicity of the smoothing parameter, $\eta_{\epsilon/n}(\mathbb Z)\le\eta_\zeta(\mathbb Z)$. Thus, all three conditions are satisfied whenever
\begin{align}
\sigma\ge 1.1658\sqrt q \eta_\zeta(\mathbb Z),
\qquad \zeta=2^{-\lambda}/4n.
\end{align}

Next, recall that previous work provides the optimal Gram-Schmidt norm for $B$ to be around $1.1658\sqrt{q}$. Therefore, we'll fix:
\begin{align}\label{eq:sigfg}
    \sigma_{f, g} \coloneqq 1.1658\sqrt{q/2n},
\end{align}
to ensure $B$ is selected optimally \cite{Ducas13, Ducas14}. Note that this also results in the average norm of the vector $(f, g)$ to be approximately $1.1658\sqrt{q}$.

Next, having fixed $\sigma_{f, g}$ and bounded $\sigma$, the only remaining parameters for the hardness of the $\NTRU$ learning problem underlying $\FALCON$ are $q, n$. To this end we'll require:
\begin{align}
    \Adv_{\sigma_{f, g}}^{\NTRU}(\calN) \le \frac{2^{-c}2^2}{(2^{c+2}+1)^2} \le 2^{-3c-6}.
\end{align}
Recall that $p_{GS} \approx 1/4$ for our parameter regime. And in order to understand block-size $\mu$ and the resulting bit security $b$, we apply the Core-SVP model and take the log to find $\mu$ for our $\NTRU$ problem such that:
\begin{align}
    0.265\mu \ge 3c+6.
\end{align}
Given that $\FALCON^{(+)}$ is originally specified as having $q$ on the order of $10^5$ and requires that $q \equiv 1\mod 2n$ for $q$ prime and $n$ a power of $2$, the hardness of this problem is difficult to tune as there are only so many pairs of $(q, n)$ in this regime that meet all of the desired criteria \cite{FALCON}. 

Finally, along with the parameters already discussed, $\SIS$ introduces $\beta$ using $\tau_0$ a tail-cut bound related as:
\begin{align}
    \beta \coloneqq \lfloor\tau_0 \sigma \sqrt{2n}\rfloor,
\end{align}
where we will lower-bound $\sigma$ as described previously. and from the third term, we require:
\begin{align}
    \Adv^{\SIS}_{n, 2n, \beta}(\calC) \le \frac{2^{-c}2^2}{(2^{c+2}+1)^2} \le 2^{-3c-6}.
\end{align}
Next, applying the Core-SVP model and taking the log, we now may write:
\begin{align}
    0.265\mu \ge 3c+6,
\end{align}
for $\mu$ the $\BKZ$ block-size needed to solve our $\SIS$ instance. The hardness of our $\SIS$ instance can be increased slightly by lowering the value of $\tau_0$ and thus $\beta$. However, the extent to which this can be fine-tuned is limited as the result is not especially sensitive to changes in $\tau_0$. 

We calculate the public key size in bytes as $\lceil n \log(q)/8\rceil$ and the signature size in bytes as $\lceil k + n\log(q)/8\rceil$. Note that as signatures include small polynomials $s_2$ rather than any polynomial in $R_{q, n}$ it is practical to utilize some form of compression and padding algorithm when storing signatures \cite{FALCON}. However, as we do not discuss this sub-algorithm in our analysis, we will not apply such compression processes to our calculation of signature sizes in the table below. We use this same method of size calculation for both the parameters found in \cite{FALCON} as well as our own work.


Combining the above requirements we present two parameter sets below and provide new security calculations for the parameters initially recommended by $\FALCON^{(+)}$'s specifications \cite{FALCON}. No parameter sets matching SL2 or SL3 are given due to the challenges in fine-tuning $\NTRU$ and $\SIS$ within the desired parameter regime.

\begin{table}[H]
    \centering
    \scalebox{1.0}{
    \renewcommand*{\arraystretch}{1.1}
    \begin{tabular}{|c|c|c|c|c|}
        \hline
        &\multicolumn{2}{c|}{$\FALCON$ \cite{FALCON}} &\multicolumn{2}{c|}{Our work} \\
        &SL1 &SL4/5 &SL1 &SL4/5\\
        \hhline{=|=|=|=|=|}
        $q$ &12289 &12289 &12289 &40961 \\
        $n$ &512 &1024 &2048 &4096 \\
        $\tau_0$ &1.1 &1.1 &1.1 &1.1 \\
        \hline
        $\beta$ &5834 &8382 &70740 &254262\\
        $\sigma$ &165.74 &168.39 &1004.84 &2554.84 \\
        $\sigma_{min}$ &1001.16 &1396.20 &1004.84 &2554.84 \\
        $\sigma_{f, g}$ &4.04 &2.86 &2.02 &2.61 \\
        \hline
        $pk$ size (bytes)       &870 &1739 &3478 &7845  \\
        $\sigma$ size (bytes)   &910 &1779 &3518 &7885 \\
        \hline
        $\NTRU$ $\BKZ$ Block-Size  &154 &421 &1024 &2170 \\
        Quantum Core-SVP            &11 &34 &87 &187  \\
        \hline
        $\SIS$ $\BKZ$ Block-Size     &234 &539 &1356 &2884  \\
        Quantum Core-SVP            &18 &45 &112 &249   \\
        \hline
    \end{tabular}}
    \caption{We give parameter sets targeting the intended security levels SL1 and SL4/5, together with concrete security estimates supported by our security proof. We also apply our analysis to the parameter sets proposed in the $\FALCON^{(+)}$ specifications \cite{FALCON}. For each parameter set, we report both the chosen Gaussian parameter $\sigma$ and the minimum value $\sigma_{\min}$ required by our analysis. Note that, unlike the original specifications, we do not account for signature compression when calculating the signature sizes reported above.}
    \label{tab:compare_fal_specs}
\end{table}
\vspace{-10ex}
\subsection{Discussion}
After concerns were raised for the security proof of $\FALCON$ as provided in \cite{FALCON}, we provide the first proof of the $\sUFCMA$ security of $\FALCON^{(+)}$ in the QROM. Specifically, we give a rigorous reduction from the $\NTRU$ learning problem and $\RISIS$ to the $\sUFCMA$ security experiment for $\FALCON^{(+)}$. We then quantify the bit security of $\NTRU$ and $\SIS$ against the current state-of-the-art attacks using the BKZ algorithm. This allows us to calculate parameter sets which can -under our reduction- ensure the security requested by NIST's PQC Standardization Project. 

Our reduction does not identify an attack on current parameters. However, we were able to verify that the largest factor in our security guarantee (Equation \ref{eq:theo}) -$O(Q^2)$, which results from the use of the Measure \& Reprogram algorithm in Lemma \ref{lem:tight}- is tight for Hash-and-Sign signature schemes (see Section \ref{sec:tightness}). This does not formally prove the tightness of our reduction overall, but it does strongly support that conclusion.  Our results invite further work on the tightness of the Measure \& Reprogram approach to proving security of $\FALCON^{(+)}$.

\subsubsection{\ackname}
\begin{credits}
This work was supported by the National Institute of Standards and Technology (NIST) and the Joint Center for Quantum Information and Computer Science (QuICS) at the University of Maryland. This research paper is not subject to copyright in the United States.  The opinions, findings, and conclusions in the paper are those of the authors and do not necessarily reflect the views or policies of NIST or the United States Government.

\noindent \textbf{Use of Generative AI:} Large language models were used in the preparation of this document to identify typographical errors, improve the readability of prose, review mathematical reasoning, and suggest potentially relevant literature references as a supplement to a manually conducted literature review. All suggested references were independently verified by the authors, who retain full responsibility for the accuracy and content of the manuscript.
\end{credits}

\printbibliography
\end{document}